\documentclass[dvips]{arxstspdf}
\usepackage{graphicx}
\usepackage{flushend}
\usepackage{stfloats}

\volume{26}
\issue{3}
\pubyear{2011}
\firstpage{440}
\lastpage{469}
\doi{10.1214/10-STS324}

\begin{document}
\begin{frontmatter}

\title{A Conversation with David R. Brillinger}
\runtitle{A Conversation with D.~R. Brillinger}

\begin{aug}
\author{\fnms{Victor M.} \snm{Panaretos}\corref{}\ead[label=e1]{victor.panaretos@epfl.ch}}
\runauthor{V. M. Panaretos}

\affiliation{Ecole Polytechnique F\'ed\'erale de Lausanne}

\address{Victor M. Panaretos is Assistant Professor of Mathematical Statistics,
Institut de Math\'ematiques, Ecole Polytechnique F\'ed\'erale de Lausanne, EPFL-IMA-SMAT
Station 8, 1015 Switzerland \printead{e1}.}

\end{aug}

\begin{abstract}
David Ross Brillinger was born on the 27th of October 1937, in Toronto, Canada.
In 1955, he entered the University of Toronto, graduating with a B.A. with
Honours in Pure Mathematics in 1959, while also serving as a Lieutenant in the
Royal Canadian Naval Reserve. He was one of the five winners of the Putnam
mathematical competition in 1958. He then went on to obtain his M.A. and Ph.D.
in Mathematics at Princeton University, in 1960 and 1961, the latter under the
guidance of John W. Tukey. During the period 1962--1964 he held halftime
appointments as a Lecturer in Mathematics at Princeton, and a Member of
Technical Staff at Bell Telephone Laboratories, Murray Hill, New Jersey. In 1964,
he was appointed Lecturer and, two years later, Reader in Statistics at the London
School of Economics. After spending a sabbatical year at Berkeley in 1967--1968, he
returned to become Professor of Statistics in 1970, and has been there ever
since. During his 40 years (and counting) as a faculty member at Berkeley, he has
supervised 40 doctoral theses. He has a record of academic and professional
service and has received a number of honors and awards.
\end{abstract}


\end{frontmatter}

\begin{figure*}

\includegraphics{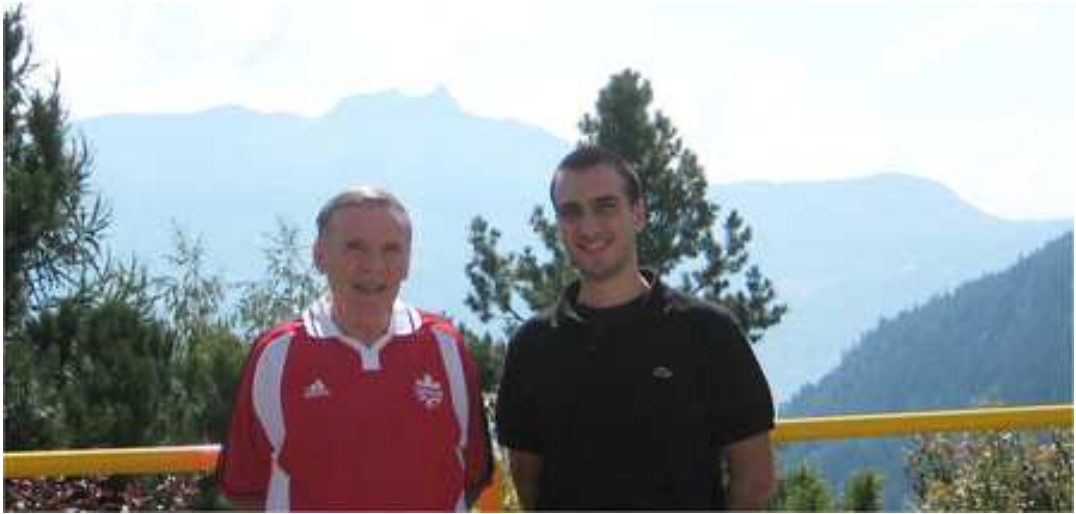}

  \caption{David and Victor with the Swiss Alps in the background. Photo taken
during the interview session, September 2009. David is proudly wearing the
Canadian Soccer team shirt.}  \label{brill_vic_alps}
\vspace*{-6pt}
\end{figure*}

This conversation took place on September 9th 2009, in the Swiss Alps of Valais,
during David's visit to give a doctoral course on ``Modeling Random
Trajectories'' in the Swiss Doctoral School in Statistics and Applied
Probability (see Figure~\ref{brill_vic_alps}).

\section{Growing Up in Toronto}

\textbf{Victor:} I suppose this is an interesting setting to be doing
this, as one story would suggest you originally come not from very far from
here$\ldots.$

\textbf{David:} Indeed! Now I don't know the specifics, but there were
Brillingers in Basel at the end of 1400s. Once we were in Zurich, at Peter
Buhlmann's invitation, and we saw a statue that was close: B-U-L-L-I-N-G-E-R. Now,
the Brillingers in Basel became protestant at the time of Martin Luther. The
next time I find them is in the 1700s when Brillingers went to Pennsylvania as
Mennonites. They finally got up to Canada after the American Revolution. They
were the original draft dodgers. You see then, in America, men had to be in the
militia, but the~Bril\-lingers were pacifists. So they went to Ontario where they
could practice their religion as they wished. So I'd like to think that there is
some Swiss background and presumably it would have been through some great--great
uncle who was ``Rektor'' of the University of Basel.

\begin{figure*}[b]

\vspace*{-3pt}
\includegraphics{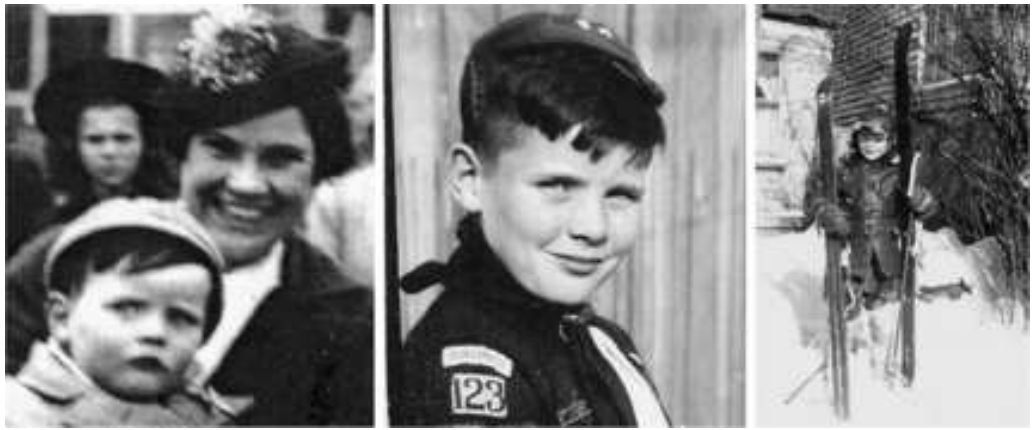}

  \caption{Young David in his mother's arms at the King and Queen's visit to
Toronto, as a Cub Scout, and with his ski gear.}
  \label{brill_young}
\end{figure*}

\textbf{Victor:} I see, I~see, so it would then be Brillinger
(German pronunciation) rather than Brillinger\break (French pronunciation)?

\textbf{David}: That's right. And you Victor told me that you've seen a
truck on the Swiss highway with Brillin\-ger on it. Also Alessandro (Villa) told
me he saw a~mailbox with Brillinger on it, or something like~that.

\textbf{Victor:} Jumping much further into the future: you grew up in
Canada.

\textbf{David:} Yes!

\textbf{Victor:} Could you tell us a bit about your family?

\textbf{David:} My father died---let's just work it out---when I was 7
months old, so this was very harsh on my mother. She woke up in the middle of
the night and he seemed to be\vadjust{\goodbreak} in some trouble, but then she fell back asleep
and I think she felt guilty about that ever after. I~doubt there was anything
that could have been done back then because he died of a cerebral hemorrhage. I~wish I
could have gotten to know them together better. You know, they had their
house, a cottage, a dog and so on.  They had a Harley motorcycle and went off on
that on their honeymoon, they had a sailing canoe$\ldots.$ Lakes and Canadian things
were very much part of their lives. My mother was actually a very beautiful
woman, when you see the pictures, with smiles (Figure~\ref{brill_young}). But the smiles
mostly disappeared after my father's death. Then, it was World War II times and most of
the men were gone. It's hard for me to imagine she wouldn't have remarried. But
it just never happened.\vadjust{\goodbreak}

She really cared a great deal about my education and structured things so that I
got a fine education. At the start, there was a bit of money---because my father
was going to be an actuary, so she had some insurance money. I~went to a private
boys' school in Toronto until the money ran out. Then, there was this school for
bright kids in Toronto, the University of Toronto Schools (UTS). I~took the exam
and got into it. UTS was very important for me. I~should mention that my
maternal grandmother was also very important, and perhaps she raised me.
She had had her husband die in the great flu epidemic and found herself with
five children to raise. So I had, I~think, a beginning that made me appreciate
being alive and not really expecting too much to come from it. I~really have
been pretty content and nonaggressive about things in my life and feel very
lucky. You know, all four of my uncles---and I've decided they were my role
models---were taxi cab drivers at some point in their lives. The way they could
just talk to anybody and the way they engaged people to some extent formulated
the way I have become. I~had a lot of paying jobs as I was growing up, including
caddying, delivering prescriptions, salesperson in a small shop.

I had a lot of cousins that were important to me because I didn't have
siblings. And there were a~lot of wonderful mother's side family gatherings. So,
I~don't think I really thought about not having a~father when young, but I do
wish I could have asked my father certain questions since we did not have much
contact with the Brillinger side of the family. That was a shame.

\textbf{Victor:} Did you have any influential teachers at school?

\textbf{David:} Oh, yes! There is one very influential tea\-cher who
taught me when I was at Upper Canada College---that was the private boy's
school. I~had not started the year there and when I transferred, he found out
that I was not very good at fractions. So, he spent some time tutoring me. Now he
was also an important person in Ontario hockey. And after tutoring me he came in
the class one day and said he had 5 hockey rulebooks and he was going to give
one of them to whoever answered a mathematical problem first. So first question,
my hand went up, one rulebook; second question, second rulebook; third question,
third rulebook! So he said, ``David that's it, you can't get anymore of those!'' I
really learned I was good at sports. Or no, actually, I~wasn't good at sports,
I~was good at math, but I was very motivated when it came to sports
(laughs)! The teacher's name was H. Earl Elliott.

\textbf{Victor:} And those were the same rulebook?

\begin{figure}

\includegraphics{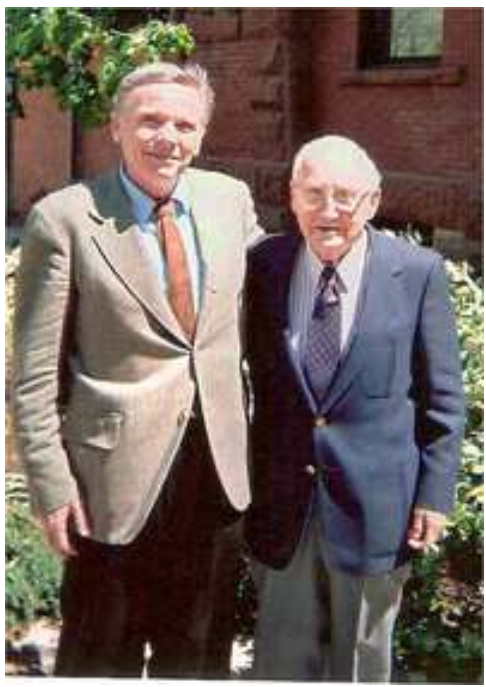}

\caption{David with Bruce ``Nails'' McLean.}\label{fig3}
\end{figure}

\textbf{David:} (laughs) Oh yes! I don't know what I was going to do
with all of them! He had not specified any rules, so I had three and gave my
cousins two! I had realized I was good at math, and I loved working on math
problems. A lot of books had problems without the solutions in the back. I~had a
lot of fun doing them. Perhaps I had more time to do that because the
weather was bad in the winter and I did not have siblings. Afterward, I~went to
UTS. I~said that was for bright kids, but part of the definition of  ``bright
kids'' then was being male (both laugh)$\ldots.$ Luckily things changed, although UTS
no longer wins the Toronto high school hockey championship like it used to! I
had a very influential mathematics teacher there, Bruce McLean\vadjust{\goodbreak} (Figure~\ref{fig3}).
He was also the
hockey coach and is still alive. He would just let me work at the back of the
room on my own. Everybody else was up toward the front, but he would
just leave me alone at this table and bring these books full of problems
(e.g., \cite{loney}). Statistics was one of the topics. And there were these British
problems that you've probably seen in the Tripos, Victor, things like that. I~don't
know about what level I~would have been at had I been in England, because
students there started working with these concepts very early on. I~read a book
where I think Dyson said he had solved all the problems in Piaggio's
differential equation book (Piaggio, \citeyear{piaggio}), but when he was at public school---I
did that when I got to University, so I guess I was lagging behind. But I think
I was very independently driven to work on these things. I~thought I solved
them, but, you know, I~didn't quite know; but anyway, I~solved them to my
satisfaction. Then, Ontario used to have some pretty tough High School exams, for
the last  year---grade 13---and four of them were on algebra, geometry,
trigonometry and problems respectively. I~got 100, 99 and 100 on the first
three and 96 on the last. I~still think about that 96. You see you were to do 10
problems, but there were 12. So I ``solved'' all 12. Later ``Mr.'' McLean told me
that the person who was grading kept getting a total of 116 on my exam, and he
could not figure out what was going on for a~while.\vadjust{\goodbreak} Eventually, he realized that
I had attempted all 12. My error was that one of them was finding the maximum or
minimum of something, so to show off I used calculus, but I forgot about
checking the second derivative! I've never forgotten that since (both laugh)!
But anyway, that brought me a scholarship that helped me make my way at
University. Back then, prizes were important because there weren't many
bursaries. Now, in America, they've switched to means tests. But I won a lot of
prizes as an undergraduate which kept my mother and me with food and so on.

\textbf{Victor:} Evidently, mathematics was one thing you enjoyed, but
what about sports?

\textbf{David:} I love sports, I~always have and I have always been a
Toronto Maple Leafs fan. I~don't know if I still have it, but there was a
wonderful picture of me about 3 years old with hockey stick in hand and skates
on feet. I~was often the last guy to make the team or the first guy not to make
the team---but I was always there! When I was growing up, they would flood the
whole neighborhood park so there would be 5 or more hockey games going on. You
didn't need all this fancy equipment. I~guess I could make the formal teams
until I was 13 or so, but then that stopped. It returned for a while when I went
to Princeton as a graduate student. There I got to be like an intramural star,
because I could raise the puck, knew the rules and played left-handed. Now,
I~mentioned my high school teacher, Bruce McLean. There's a story I love
concerning him: there was my 50th High School reunion a couple of years back and
I was in Edmonton the week before the reunion and was going to need to be in
Toronto the week after, so it was just too much time to be away from Berkeley.
One of my dear friends from High School and University, John Gardner
(now Chair of the Board of Directors of the Fields Mathematical Institute), asked if I'd
like him to arrange a lunch with ``Nails'' McLean---his nickname for UTS
students was ``Nails.'' I said of course! So, when I went to Toronto the week
after, we had lunch. McLean was 96, and had driven in through all the traffic to
central Toronto for the lunch. We had a wonderful time. It turned out he had
also been in the Navy, so we discussed that. But at the end of the meal he got
this incredibly serious look on his face. So I'm thinking, ``What's this all
about?'' And he says ``David, when you were at school, there was something I
really worried about, I~worried about it for a long time.'' So I'm sitting there
with my eyes rolled back and wondering. He continued, ``I really wanted you on
the hockey\vadjust{\goodbreak} team, but there were a lot of good players that year!'' (both laugh).
I~just grin when I remember that. And indeed the team was good. They won the
Toronto championship. I~just wanted to get the sweater, go to practice, and, if
we're winning 7--2, get to skate around a bit. But I had to wait until Princeton
to do that.

\section{University of Toronto and the Canadian Navy}

\begin{figure*}

\includegraphics{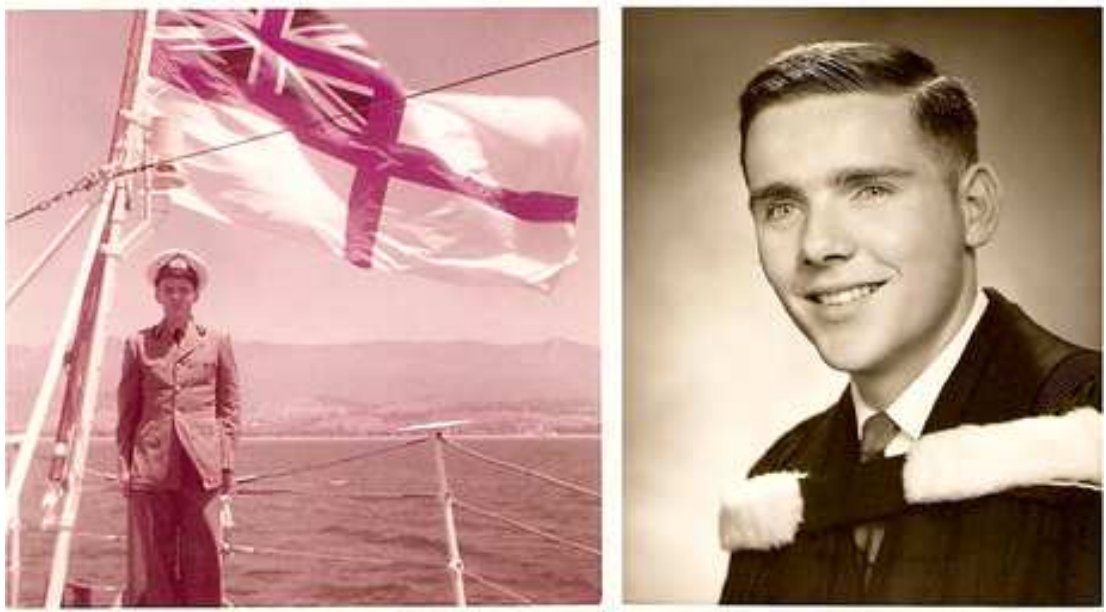}

\caption{David in the Navy off Santa Barbara in 1957, and upon graduation from the University of Toronto in 1959.}
\label{brill_navy}
\vspace*{-3pt}
\end{figure*}

\textbf{Victor:} You mentioned before that you were in the Navy, can
you tell us a bit more about that?

\textbf{David:} That was at University. I~knew that by joining the
Navy I was going to get to go outside of Toronto and perhaps Canada for a bit;
because Toronto was really a bit boring back then. Canada did not have a draft---still
doesn't---so the way the government thought they could get officers for
the regular military was by having army, navy and air force programs at the
universities. That was a bit like Boy Scouts, and I'd been a Cub (Figure~\ref{brill_young})
and a Boy Scout. For me, it was obvious to join the Navy because I loved to
canoe and sail, and you got to go to Europe and Mexico. Whereas if you were in the
Army, you got to march around in the dust of Ontario; and if you were in the Air
force, you were in Saskatchewan, which is flat, and with not so much to do then.
So, I~was on my way to seeing the world and at the same time got paid very well;
the food and the clothing were obviously provided. Plus, it was a lot of fun, I~just
loved it. I~mean guns were only 5\% or less of the life. So it was a
no-brainer to be in the Navy. Second year I was based on the West Coast (Figure~\ref{brill_navy}).
In the program there was a prize for the person who was best in navigation and I
think I won probably easily, as I had taken an astronomy course and had learned
all this spherical trigonometry previously. The way things worked, I~ended up
being a communications officer learning about radio and coding. This was great
since I had been learning physics as well as mathematics. You know, in my career
I've gotten to study mostly the things I was good at and enjoyed. I~was
principally good at math, and it was obvious what my career was to be.

\textbf{Victor:} You once told me a story about doing some very applied
statistics in the Navy.

\textbf{David:} That was my first independent statistical research
activity, I~would say! So let's think. My fourth summer, I~had already gone
through a lot of basic training, becoming a communications specialist and a sub
lieutenant. I~was going to be in the aircraft carrier, the Bonnaventure, and we
were supposed to sail into the middle of the Atlantic because the Queen was
going to fly over there on her way to visit Canada. And so we were to be
stationed out there. I~don't know why, maybe in case she leapt out with a
parachute or something like that! I mean it was awfully ill-defined
(both laugh)!

\textbf{Victor:} $\ldots$after all it is the \textit{Royal} Canadian Navy!

\textbf{David:} Exactly! So we had to toast to the Queen~at banquets
and such and such. Anyway, they had to find something for me to do during the
open period before the mission. So, they decided that, since I~was studying
statistics, they would like to know how~ma\-ny messages were sent out by the fleet
weekly for~se\-veral years. They took me to this room, and here~we\-re these huge
stacks of signals by week. I~would still be counting them if I had done it
directly! But instead I thought why don't I just get 100 and weigh them and
estimate a weight per signal. And then~I~as\-ked for a scale, which they found.
And I just measu\-red how heavy the piles were, and so I gave them~nice graphs.
When the fleet was at sea, there were~a~lot more signals, and things like that.
I~guess it sounds nutty to be saying the following, I~mean I'm totally a pacifist
and I think I've been that all my life---but I did enjoy the Navy! I suppose
back then Canada was doing peace keeping. Like Brazil's these days, that was the
Canadian role then. Our Prime Minister Lester Pearson won the Nobel peace prize
for the idea of creating a UN Peace Force. My thought was that the world needs
policemen, and since Canada~was not in an aggressive posture at that point,
I~signed up. By the way, in the remaining time before the~crui\-se, I~did a lot of
dinghy sailing in Halifax harbor.

\textbf{Victor:} Shall we talk a bit about the University of Toronto (U
of T)? You did your bachelors honours in pure mathematics. I~recall you telling
me in Berkeley that you were already reading Bourbaki as a first year
undergraduate---in French.

\textbf{David:} Yes, that's true! I was lucky because Cana\-da was
trying to be bilingual to support its francophones  and I studied French for
seven years. So there was a professor at U of T, John Coleman---who is still
alive, aged a hundred or so I think; these Canadian mathematicians live a long
time. He found out I could learn and read in French. I~think he identified me
especially because I had won this prize for algebra/geometry/trigonometry and
problems. He found what I looked like by
watching where my homework handed back ended up in the classroom. He invited me
for a coffee or whatever. Actually, he was remembering when I talked to him a
couple of years ago that we had butter tarts and tea when we met. He got me
reading Bourbaki. And then he said why don't you do some of these problems? So
we met then each week: I~couldn't do the problems, and perhaps he had trouble
too. I~don't know if I could do them now, it would be fun to try. The first
book was on algebra and I believe that Coleman bought it for me. I~still have it
(Bourbaki, \citeyear{bourbaki}). The later ones on analysis have probably been the most
important to me. Coleman got me reading Bourbaki and I remain very appreciative.
Going through them really stood me in good stead when I got to Princeton. I~found myself
a couple of years ahead of the American students. You see I'd
gotten to do mainly maths and physics at Toronto, and I also had this secret
weapon: French! I mean the French probabilists were then doing all this
wonderful stuff, E.~Borel, P.~L\'evy and M. Fr\'echet, for example. And most of
their things were not being translated. Nowadays the French mathematicians write
in English most of the time so that's not an issue. That was first year. That
year I also had a course from Ralph Wormleighton, he had been at Princeton---there was a
real Toronto--Princeton railroad including Don Fraser, Art Dempster,
Ralph Wormleighton; and when I applied to grad school I only applied to
Princeton. It never occurred to me to apply anywhere else. I~don't think that
was a statement of confidence, but I didn't have anyone who had been at
university at home, so I just was not getting that kind of advice. The second
year was Dempster. Dempster has often taken the geometric approach. When I took a
course from Coxeter, I~later saw where that approach was coming from. And then in
the third year was Don Fraser---he was certainly using a lot of  algebra. The
fourth year was Dan DeLury. He was this skeptical older guy. He'd been out doing
biometrical studies. His attitude was that one might have thought that they had
designed an experiment well, but there were many ways that an experiment might
have gone wrong. His course was very maturing for me. It's important to have
some training in criticism when you're an applied statistician.

\textbf{Victor:} So, that means that you would have had quite a
rigorous maths background but also would have been exposed to quite a bit of
statistics, which is rather atypical for that time period.

\textbf{David:} Although I was in pure mathematics---\break that's~what my
degree was in---I went to all the statistics courses. As a matter of fact, I~probably
went to \textit{all} the courses, including the actuarial ones. Back then, I~could just
sit there and absorb things. It's not as though I'm boasting; I
used to feel embarrassed about saying things like that, but I think I was just
lucky: it was not really anything I did, it's just the way it was. I~wish I
could have played hockey better, but I didn't get that skill nor the ability to
run 100 meters in less than 10 seconds. I~guess I'm saying there may be a gene
that I was lucky enough to\vadjust{\goodbreak} get.

\textbf{Victor:} Do you recall any lectures that you particularly
enjoyed? Coxeter had a fine reputation as a lecturer I suppose.

\begin{figure}

\includegraphics{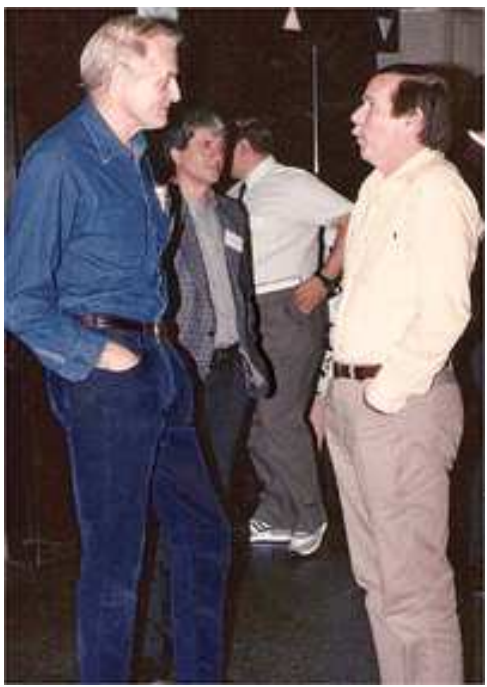}

\caption{David with Don Fraser.}\label{fig5}
\end{figure}

\textbf{David:} Oh yes, Coxeter was wonderful. He had left England
after World War II. Also Tutte, who is another geometer, was great. In fact,
Tutte had broken one of the important Nazi codes in World War~II---and none of
us knew that. But some people in the class were mean to him because he was
a~little shy, and they teased him. I'm sure if they had known about his breaking
the code, they would have been more like ``wow'' instead. Regarding Coxeter, I~remember
one funny story, where he was talking about a particular geometry for
many classes. His course became his book (Coxeter, \citeyear{coxeter}) or the book was part
of his course. So, there was this particular finite geometry he was talking
about a lot, with very bare assumptions and he was talking about it during a
number of classes. So, finally, I~asked,``Why are you spending so much time on
this, is it \textit{that} important?'' And he said something like: ``Well you
seemed so interested, Mr. Brillinger!'' I mean, I~was just asking questions to
keep up with where he was going! I was intending to become an actuary for many
years, in part because my father worked for Imperial Life. And they were very
good to my mother and me. I~had realized that if you are poor but good at
mathematics, then an actuarial career was a route to the middle class. I'm not
sure I was after being middle class, but I needed to help my mother, so I was
going to be an actuary. But Don Fraser, who had great influence on me
(see Figure~\ref{fig5}), 
said something like: ``Well, David, sure that's nice, that you're going to be an
actuary, but why don't you go to Princeton first?'' So, I~did! I went to
Princeton, the plan being to become an actuary after I was done with all this
childish fun, namely, mathematics.\looseness=1

\textbf{Victor:} Apparently it was \textit{too} much fun$\ldots$!

\textbf{David:} I guess that's right. And I realized at some point that
anything I could do as an actuary, I~could probably do as a statistician---with
the added benefit that I would get to travel and be an academic. I~did take
enough of the exams to become an Associate of the Society of Actuaries.

\textbf{Victor:} Just before going off to Princeton, you were among the
winning five of the Putnam competition of Spring'58.

\textbf{David:} It was again Coleman who got me involved.

\textbf{Victor:} And I recognized a couple of other famous names on the
same honours list, Richard Dudley and Larry Shepp.

\textbf{David:}  Yes, I~got to know them both. You see, both of them
went to Princeton for graduate studies. I~really had no idea of what was
involved. I~just went and took the exam! I remember that Erd\"{o}s visited
Toronto for a month and he gave a course. One of the problems he taught us was
on the Putnam exam! (laughs) Some number theory thing (continues laughing)$\ldots.$ So
on the exam day that one was out of the way pretty quickly! He was just a real
gem, a real role model. I~mean he had these simple direct ways to approach
problems, and would advocate that you should take a breath before you start
writing down a lot of equations and things like that. U of T was absolutely
super. I~got a super education in mathematics there and at high school. I~mean
some people might think of Canada as being a backwater, or as having
been one, but there were some very fine researchers and teachers. You know,
Coleman had also gone to Princeton just before the War started. I~was
lucky.

I can't resist adding that, while I was at U of T, I~was actually at Victoria
University. There, I~earned a letter for playing on the soccer and squash teams,
each for four years. I~can show the letter to you!
I would also like to add that Art Dempster and Don Fraser
have long been role models for their ex-students. In research they each have taken roads less travelled
in their work.

\section{Princeton}

\textbf{Victor:}  When did you move to New Jersey?

\textbf{David:}  In the summer of '59. That was my last summer in the
Navy, and I had become a Lieutenant. I~turned up there in the beginning of
August having left the Bonnaventure. I~had asked if there was some work for
me,
and it turned out that Sam Wilks had just finished writing his book Mathematical
Statistics (Wilks, \citeyear{wilks}). My job was to work on the problems. I~remember I just
lay out under the trees at Graduate School working on them, right by the golf
course---which I~would golf on most days, illegally. I~remember going over to
Wilks' office just before term started. One of my Canadian friends, Irwin
Guttman, was there. I~said, ``Well here are the solutions, but I couldn't get one
of them.'' And Wilks went ``What???'' In the end he took that problem out of the
book. It was about proving that the median and the mean were jointly
asymptotically normal. It took me a while to figure out a neat way to do that.

\textbf{Victor:}  You got right into mathematical statistics upon
arriving at Princeton.

\textbf{David:}  Oh yes. Already at Toronto, I~could see that
statistics, perhaps as an actuary, was for me, because you interact with people
a lot. Math was a~lot of fun too, but you interact with a much narrower group of
people. DeLury had impressed me, because he was really working at the
frontier of the applications of statistics. I~have found myself realizing that
statisticians are the keepers of the scientific method. When a scientist
comes up with something, what can they reasonably conclude? That appealed to me,
to be able to get involved in many fields.

\textbf{Victor:}  And when did you meet Tukey?

\textbf{David:}  (laughs) Aaaaah,  John Tukey$\ldots.$ I watched him like a
hawk! Because he was so interesting generally and so much fun to watch. I~had
been told about Tukey by Coleman. Coleman had been a graduate student when Tukey
was at Princeton. And Coleman told me that I was going to meet someone who, at
beer parties, was always drinking milk, he just had a big glass of milk. So
I knew before meeting him that Tukey was different. Because at a beer party in
Canada you drink beer, that's part of your manhood, or something like that.
Princeton; at Princeton you didn't have to take any courses. You could sign up
for one and would get an A, even if you never turned up. You had to write a
thesis and pass an oral exam, so that was pretty good! So let's see; Tukey
gave a time series course. And here was this person, unlike any other person I
had ever met. He was from New England, very Canadian in a lot of ways. He had
pride in his background. He was careful with money, and he had apple pie for
breakfast. So I~went to his time series course and this involved a
lot of Fourier analysis---and I had a strong background in trigonometry and
that made the course attractive.

\textbf{Victor:}  Did you attend any of these courses along with David
Freedman?

\textbf{David:}  Oh yes! David F. was a year ahead of me, and he
was influential on me (pauses and reflects for a moment). I~guess, oh my, most
of these people are dead now, goodness. OK, whatever. I~have these two stories
about David, one involving Frank Anscombe and the other John Tukey. Now, David
was a year ahead of me at Princeton. He was from Montreal, I~was from Toronto so
we were natural ``rivals,'' right from the beginning! That's just the way it was.
Of course I don't mean that in a bad way. Anyway, Frank had asked David
F. to be his teaching assistant in a course. And David said, ``but I am on a
scholarship, I~don't have to do that!'' ``OK, fine,'' said Frank, and then Frank
asked me (laughs). And I knew what David had said, and got to give the same
answer! David analyzed a lot of situations very clearly, and I observed David as
I do a lot of people.

David F. never changed in terms of his intellectual calibre and wit, and
the character of his questions. David was also in Tukey's time series course.
Early in the term Tukey used the word spectrum several times. And David after, I~don't know,
20 minutes or some such, asked what the definition of a spectrum
was. So, Tukey said something like: ``Well, suppose you've got a radar
transmitting signals up and it bounces off an airplane and a signal returns $\ldots$
so you see $\ldots$ well that's a spectrum.'' So, David's manner was ``Well, ok.'' Then
the next class the same thing happened. Tukey mentioned the spectrum, David
wanted a definition, and Tukey said, ``Well, suppose you have a sonar system and
it bounces a signal off a submarine, or some such''$\ldots$ David never came back
(both laugh)!

That was really pure David F., wanting clear explicit definitions. Tukey
and David were the opposites of each other. You see, Tukey believed in
vague concepts. He believed that if you tried to define something too precisely,
then you would have lost important aspects going along with it. But David didn't
think that you could talk about things properly unless you were completely
clear. Of course, Tukey's and David's great confrontation was over census
adjustment. I~picture that David took a strict interpretation over what was
required, while JWT was after an effective estimate of the counts. It is no
surprise that David was debating champion at McGill. He surely could have
been a fine lawyer, and then a judge, and then$\ldots.$

\begin{figure*}

\includegraphics{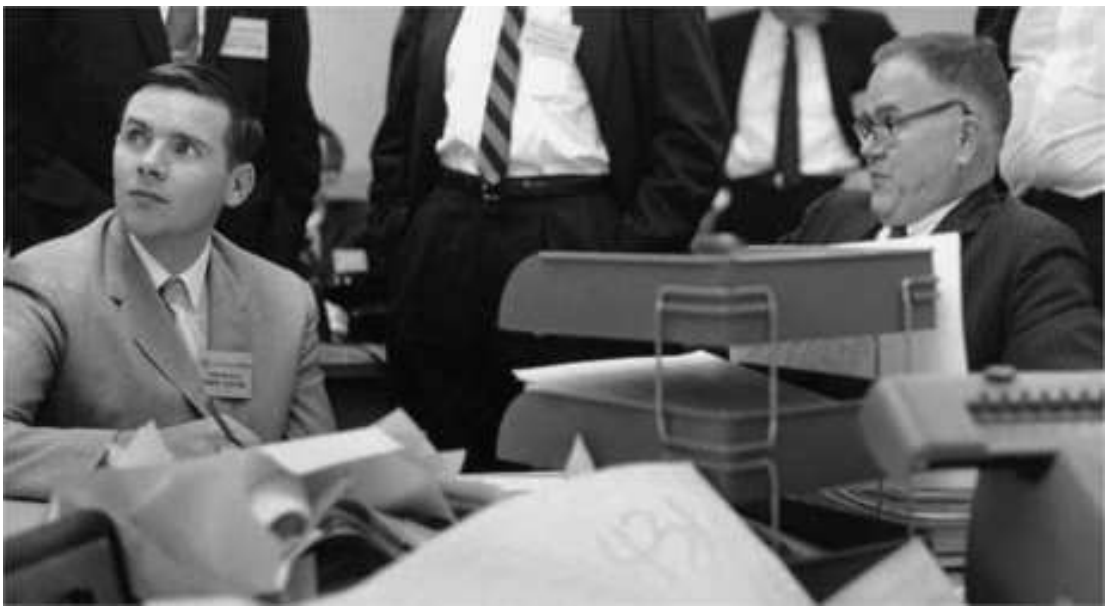}

  \caption{David with John Tukey at the NBC Election Centre in
  1962.}\label{fig6}
  \end{figure*}

\textbf{Victor:}  He did get involved with statistics and the law.

\textbf{David:}  Yes, he was involved  in statistics and economics,
too. He worked at the Bank of Canada for a~while. I~think he might have expected
that he would be going down that road. He probably thought that being~a
statistician you can do anything you want to---that was my own reason for
choosing statistics.

David was a very sweet person. I~am thinking just now of his taking Lorie and me
out to dinner in a~nice Princeton restaurant after we got back from our
honeymoon.

\textbf{Victor:}  Going back to Tukey, what did you learn from him as a
researcher, what was his style?

\textbf{David:}  I learned that there are novel ways to solve most
problems. I~think JWT could add two four-digit numbers in ten different ways
that no one else in human history would ever have thought of! I mean he was like
Richard Feynman. He was of the same ilk. There are people, and there are lots of
historical examples, who just think differently than almost everyone else. Also
what I have learned from Tukey is that there is a physical interpretation of so
many of these concepts when you look at the history of mathematics. That's what
I tried to bring up in my talk this morning about how some of these things came
out of Kepler and Lagrange and so on
(David was lecturing on SDE modeling of random trajectories using potential functions).
That you can understand a lot of
this contemporary work if you think about how it had been generated in the first
place. I~think Tukey often found himself explaining things to people who didn't
know much mathematics. I~paid attention to how he did that. I~would like to
think that I'm not bad at doing that too. In a~sense, you probably lie a bit, I~mean you
probably use an analogy or a metaphor at some point, which is not quite
right, but people get the idea.

\textbf{Victor:}  That's the advantage of vagueness.

\textbf{David:}  Yes, indeed! Tukey's vagueness meant, for example,
that we could start out with standard errors and later find ourselves talking about
the interquartile range, just letting the idea of ``spread'' be vague.

\textbf{Victor:}  What was your relationship like when he became your
advisor?

\textbf{David:}  
There were lots of good problems around
Fine Hall and the Labs that I worked on. Eventually, JWT suggested a particular
one. The deal seemed to be that if I started to have trouble, I~should go see
him. Maybe his not being around town often was part of the breaks in our
meetings. When I would meet him, if I seemed a bit too cocky, he would knock me
down; and if I looked discouraged, he would build me up. My thesis
concerned formalizing Gauss's delta method by working with truncated random
variables asymptotically. Another thing was that during the school year I~had
the day-a-week job at Bell Labs, so often I drove back and forth to Bell Labs
with him, sometimes in his convertible. During those drives, we talked about a
lot of things. Sometimes, there were other passengers too. I~learned while
working with him that, when he used some new word, I~shouldn't worry about it.
I~should just let him talk a while and then try to figure out what it was all
about. I~think a lot of people had a hard time understanding what he was trying
to get at. I~would eventually come up with something; now if it's really what he
meant, I~don't know. I'd say I had a wonderful relationship with him (Figure~\ref{fig6}).
I~would kid him---I~mean I didn't know you shouldn't tease professors until much later!
Because I was working class Canadian and had my uncles as role models. That's how
they'd approach people. Not mean teasing, just seeking a smile. I~have also
teased David Cox. David was patient with me.

\textbf{Victor:} There was good chemistry between you, then. Because,
you know, he was relatively conservative and you've been pretty progressive and
open about it all along.

\textbf{David:}  There was, yes sure. We could talk about things just
like that. No tension. He was on the conservative side, true. But it was more
about different cultures. He was American and I am Canadian.  Canadians are
progressively conservative. In those days, there was a conservative spirit in
Canadians when it comes to the way one dresses or the way you talk to other
people. So, there was conservatism in me, but it was social conservatism, not
political conservatism.

\textbf{Victor:}  Well, it would appear that Tukey had a~ve\-ry high
opinion of you. It has been rumored that he used a ``milli-Brillingers'' scale
to measure people up?

\textbf{David:}  (laughs) Yes, I~have heard that from several people,
including Mike Godfrey and Bill Wil\-liams, but what does one say? Bill told me
that once Tukey asked about a prospective student, ``How ma\-ny milli-
Brillingers?'' Bill's reply was ``four or five hundred mB's.'' John responded
with something like, ``Well that's very good.''  I don't know, I~guess that I was
quick on my feet, I~don't mean at running. If I had to do something, I~would go
and do it.

\textbf{Victor:} What about Sam Wilks whom you just mentioned earlier?

\textbf{David:} Sam was wonderful too. He was just a gem. It's a shame
that he died way too soon. One story is that he was taking shingles medicine and
drank some alcohol that night and there was a bad synergy. Another is that there
was an unpleasant meeting over the admission of a student to the program. Sam
was conservative politically, but that was never an issue. He had me work on
these problems in the draft of his book as I mentioned. I~also sat in on the
course that was based on the book he was writing. He was a social animal. I~can
tell you one story. The Tukeys---God knows for what reason---had decided to
have a come-as-your-spouse party. So Lorie was supposed to dress like me and I
like Lorie, and so on and so forth, Mrs. Tukey like John Tukey, and John Tukey
like Mrs. Tukey. That happened, but Gena and Sam Wilks came along as themselves!
Near the end of my studying, I~went off for an interview at the University of
Michigan, before I~knew whether I would receive a postdoc. Jimmy Savage was
there then. I~told him about the party. And I think he went like this
(David holding his chin down) and said, ``I know too much Freud to ever do something
like that!'' I didn't know a lot about Freud and I still don't know what Savage
meant, but he did know a~great deal about a great deal of things.

\textbf{Victor:} So how did you meet Lorie?

\textbf{David:} Blind date! And we're both proud of that! One has to
take risks sometimes. She went to Antioch College with its work--study program.
She was studying sociology and had taken a statistics course using Mood and
Graybill---not an easy book. She was in Princeton in the ``work'' component at
the commercial side of the Gallup Poll. The Riehms introduced us. Carl was in
mathematics, eventually becoming a professor at McMaster University, and Elaine
was also working at Gallup. I~think her and Lorie's desks were next to each
other. The Riehms were often trying to get Lorie and me together, but Elaine
kept complaining because I was always out of town!  I went back to Toronto a
lot---no course responsibilities, remember? Lorie was attractive and we found
lots of things to talk about. Anyway, it was a blind date. And, I~don't know, we
just hit it off quickly! One thing that I loved about Lorie was that she was
very political---my politics weren't well formed at all yet---and she was also
very analytical. Her parents even more so! Later, we realized that we each had a
parent who had been born in China, the child of Methodist missionaries.

\textbf{Victor:} What a coincidence!

\textbf{David:} Oh yes! They were, in fact, in the same part of China:
Sichuan province. And now with the web, you can find surprising things. So, I~entered my
Brillinger grandfather's name and her Yard grandfather's name, into
Google, and then found them in the same book (Bondfield, \citeyear{bondfield})! Lorie's
grandfather was in an American missionary and my grandfather was a Canadian
medical missionary. Her parents were very political and they had a huge wealth
of political literature. Probably like the literature you, Victor, grew up with. I~was a
bit shy with them, and since they had all these magazines and books on
the coffee table, I~could always check something out while I was listening. So,
there was a very political side to it all, too. Anyway, we fell in love and it's
been good.  Almost 50 years now! People often say about us that we don't need to
talk, that we just simply communicate. Lorie changed her career goals quite
drastically after meeting me. If she had returned to Antioch College, then I
would have gone to Yellow Springs with her, probably to teach statistics. But in
the meantime, I~completed my Ph.D. and had applied for a post-doctoral fellowship
at London, which I was awarded. Lorie decided she preferred to go to London. She
was actually studying British Trade Unions at Oxford when I asked her to marry
me, so she got back to England quite quickly.

\textbf{Victor:} Indeed, you really dashed through your Ph.D. in less than
two years! How did that work? Did the lack of coursework requirements have
anything to do with that?

\textbf{David:} I don't think so.

\textbf{Victor:} I guess that your ``milli-Brillingers'' had!

\begin{figure}

\includegraphics{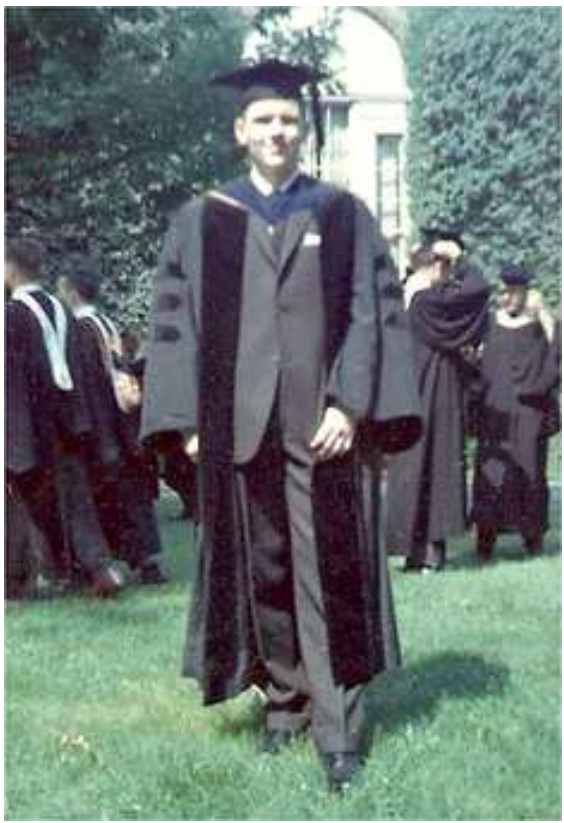}

\caption{David in his Princeton Ph.D. Regalia in 1961.}\label{brill_phd}
\end{figure}

\textbf{David:} (laughs) Aaaah, I~don't know, I~guess Tukey gave me a
problem, and said, ``see what you can do with it.'' So, I~graduated that following May
(see Figure~\ref{brill_phd}). Why
didn't he give me something like Fermat's last theorem, I~don't know! But I
actually had a try at proving that in high school. I~read a lot of the history
of mathematics.

\textbf{Victor:} I suppose nowadays in Berkeley, as well as many other
US universities, there is quite a bit of structure with a lot of coursework and
exams. How do you compare those two different systems?

\textbf{David:} Well Freedman and I talked about that once. And we
agreed that we would not have gone to Berkeley,\vadjust{\goodbreak} which is pathetic. But that's
the system. Plus, Princeton was very selective when I went there, I~think, two
statisticians admitted each year.

\textbf{Victor:} David Cox once told me that the less structured
approach is appropriate for the very brightest of students.

\textbf{David:} Yes, I~think so, but I certainly don't claim to be a
member of that group.

\textbf{Victor:} What do you think happened with the Princeton group?

\textbf{David:} From hearsay, I~think I can make a reasoned guess.
Tukey was a dominating figure. I~know he had tremendous respect for Sam
Wilks, but I'm not sure about some of the other people there. Also, he had the
mathematicians to contend with. Yet, he needed people. He asked Don Fraser
various times to go to Princeton, he asked Art Dempster various times, he asked
me several times. Clearly, I~can only speak for myself. I~just wanted to do some
things that were mine. It sounds selfish, but Tukey was so dominant and so
quick. I~don't think that he thought any less of me because I refused. A lot of
people were afraid of him. For example, if they had a cockeyed idea, he didn't
mince words. He told me once that he thought the best way to get a scientific
discussion going on something was to start\vadjust{\goodbreak} an argument. Now that's just the
reverse of my personality. I~did see him do a lot of that. It was possible he
wanted to get beyond the early pleasantries that go on. He did run over quite a
number of people. He liked to argue and expected to win. I~think that he wanted
to win because he had a goal and wanted to get there quickly. I~did love
interacting with him during my thesis research. I~found I could communicate very
easily with him. But still, I~felt a need to do my own thing. Princeton did get a
viable group at one point, and it became a department. The members included
Geoff Watson, Peter Bloomfield and Don McNeil. They each had a definite presence
in the statistics world. However, I~think that Peter Bloomfield just got fed up
with being Department Chair. So he went off to a large department at North
Carolina State. And McNeil went back to Australia. Also, I~gather that Watson was
treated quite terribly by the Mathematics Department. I~was very sad when Geoff
died for he had spoken truth many times. Eventually, Tukey was the only senior
person left and when he retired the department went away. So, it is a sad story,
but part of Princeton's strength in statistics was that the people it was
producing for many years came through mathematics, so there was no messing with
them in terms of mathematical stuff, but yet these people wanted to apply
mathematics as opposed to doing research in some mathematical specialty. To
deviate from the present topic slightly, I~have long found classical applied
mathematics a bit boring and old-fashioned, but I do know that
Fisher 
wrote that, ``Statistics is essentially a branch of Applied
Mathematics'' (Fisher, \citeyear{fisher}).
Nowadays, one might say that statistics is a combination of applied
mathematics and applied computing, the two driving the field. A Princeton review
committee was set up, and recommended against continuing the Statistics
Department, and that was that. But I did have a lot of fun at Princeton.

\section{Bell Labs}

\textbf{Victor:} Could you please tell us a bit about your summers at
Bell Labs?

\textbf{David:} The first summer in grad school, there was a group of us
from Princeton that had summer jobs at Bell Labs. I~would drive up there with
my friend Carl Riehm, an engineer and a logician. I~don't know if the Labs had
this program to find future employees or if it was just a good deed for science.
I~had learned some computing at Toronto on their IBM 650. Toronto had these
computing services very early on, for example, they had a Feranti\vadjust{\goodbreak} from the mid
'50s. So, I~had started out learning computing in a course in the physics
department. This was before Fortran existed, so we were using machine language.
Princeton had a 650 also, which I didn't really use that much---I guess I was a
lot more interested in group theory then. But when I went to my summer job at
Bell Labs, they had an IBM 701. Fortran got created and so they had me
programming various things for Tukey. That was pretty much the story during my
first summer; it was nice to make the money. Then, the second summer$\ldots.$ Let's
think$\ldots.$ I~guess the second summer Lorie had appeared on the scene! So, we had a
lot of fun. I~think that's when Tukey had me writing some programs involved in
discriminating earthquakes from underground explosions. He was then involved in
the Geneva negotiations for a nuclear test ban treaty with the Russians. Tukey
had one of those out of the box ideas, the cepstrum. He thought this might also
work for pitch detection. That's what I was doing. Specifically, taking speech
signal, digitizing it, doing things to it on the computer, then reconstituting
it and listening to it. Really, the spectrum and a lot of these time series
things had a real meaning for me at that point. I~also golfed a lot. The Labs
had a short 3 hole course.

\textbf{Victor:} You got experience with getting your hands dirty with
data.

\textbf{David:} Oh yes, right away. I~really loved that. But, more
importantly, I~got exposed to a whole cast of characters creating exploratory
data analysis! John Tukey was the leader, obviously. But there were
others right up there with him, Martin Wilk, in particular---he wrote some
important papers with John. There were also Roger Pinkam, Bill Williams my
buddy, Dick Hamming, Ram Gnanadesikan, Co\-lin Mallows who had a strong influence
on me. I~was in an office with Colin so that was enjoyable and educational. And
lunch was where I became a statistician, really. The whole group of us would go
down to the cafeteria and sit around a big circular table. So, lunch was about
this communal group trying to help each other with their scientific and
statistical problems. Then, people would go back to their offices and do their
own things. I~mean the old Bell Labs worked wonderfully and it's just pathetic
that it went away. There was an open door policy and everybody shared the
problem they were working on. We had a lot of fun playing pranks up there, too.
You know, it was all a gentler world back then in the early 60s. It had an
incredible influence on my becoming\vadjust{\goodbreak} a statistician because really they we\-re
creating a lot of applied statistics. I~was very lucky. I~mean I got onto a
pretty good escalator going up. You don't realize at that time how special it
all is scientifically and socially. When I've talked to some of the other Bell
Labs people, we've all said, ``Those were magic years,'' and that we were so lucky
to be right in the middle of them. Bell Labs was clearly years ahead of people
in digital signal processing. Tukey coming up with the Fast Fourier Transform
was just part of it. He was working on EDA methods
too$\ldots$.

\textbf{Victor:} Did you ``witness'' the FFT being developed?

\textbf{David:} Tukey's form, yes. In his time series course, John had
some
way of doing it by complex demodulation. Filtering this and filtering that and
then putting things together. But one day in '63, he tur\-ned up at a class with an
iterative algebraic approach to computing the discrete
Fourier transform for the case when one could factor the number of observations
into a product of two integers (Tukey, \citeyear{tukey}). It turned out that F.~Yates and
I.~J. Good had a~related way for getting the effects in factorial experiments.
The FFT idea switched a lot of Bell Labs effort from analogue to digital signal
processing. It was wonderful to be there. It gave me things to do in statistics.
The people involved got to be five years, maybe even more, ahead of the rest of
the world.

\vspace*{3pt}\section{London School of Economics}\vspace*{3pt}

\textbf{Victor:} How did England come about?

\textbf{David:} Well, part of the Canadian educational
per\-spective---and maybe you felt this too even though you are from Greece---was
that your education wasn't complete until you spent some time in England. It was
that simple. So, I~finished my doctorate, applied for a post-doc and got one!
And then Lorie and I were off to England and to the London School of Economics.
Actually, come to think of it, I've applied for only one job in my life that I
wasn't offered. See I've been in the Navy, and then Lorie and I met up. She had
strong political beliefs and I had strong social ones. Both of us were concerned
with doing things about poverty and helping the developing world. So, I~applied
for a job at the United Nations---they were advertising for a statistician.
Didn't even get interviewed! Didn't get it! Sometimes I think of how different
our lives would have been. It is impossible to know, but things have certainly\vadjust{\goodbreak}
worked out.

\textbf{Victor:} $\ldots$for statistics definitely, but maybe not so for the
United Nations!

\textbf{David:} $\!$(laughs) Sample surveys, I~think that's what they were
looking for.

\textbf{Victor:} But you've been involved in the International
Statistical Institute, which has this attitude of solidarity too.

\textbf{David:} Oh, yes, definitely! That's been traditional and I'm
glad I've had the chance to get involved in that. Anyway, England was about
completing my education and I guess something led me to the London School of
Economics. I~am not sure just what it was, but that was wonderful. Because
Kendall had just retired but was still around, Jim Durbin had just become a
Professor, Alan Stuart was about to become one too, Maurice Quenouille was a
Reader, Claus Moser was a Professor, as was R. G. D. Allen. I~was surrounded by
these senior people who were right in the middle of analyzing fundamental
economic and political structures. It was pretty good, exciting even. They
used to call these grants ``post-doctoral drinking fellowships'' (both laugh).
Lorie and I bought a Renault Dauphine and we went all over Europe. It was pretty
cheap and safe then. Fred Mosteller wanted to offer me a job at Harvard when I
came back, but he could never track me down. We were traveling to Austria for
skiing!

\textbf{Victor:} Was there any difficulty in adjusting to the British view on
statistics, having been raised to the American attitude?

\textbf{David:} No, not really. I~mean in Toronto then there was a very
British background culture there. Dan DeLury was a common sense person who
said once that he reread Fisher's Design of Experiments every year. I~think I
was different from the other British statisticians at the time, however, as I knew a
fair amount of mathematics. Nowadays there are a lot of British statisticians
who know a lot of mathematics. I'm afraid it sounds like I'm boasting too much
just now. I~saw Jim Durbin one time and he had some paper. He said he had tried
to figure out something in it a few times but failed. He asked me, ``David can
you explain this?'' I could tell at a glance that it was incorrect and said so. Jim
said, ``I wish I had your confidence.'' What he didn't have was my training,
that's what the difference was.

\textbf{Victor:} Did you enjoy the RSS meetings?

\textbf{David:} Very much. I~had never seen anything like them before
in my
life. There were people like Jack Good. He would stand up
and be coming from a~totally outside-the-box angle. I~respected that because I had
seen Tukey doing that all the time. At this point in my life, I~believe that I
have read most of Good's papers. I~was honored to be asked to speak
at his 65th birthday. I~paid a lot of attention to what David Cox, Maurice
Bartlett and George Barnard had to say, in particular. The way the meetings
worked back then was that people could get the galleys of a meeting's paper
before it was presented. So, you could compete with all these famous guys. You
could read the papers and see if you had something to add to the discussion.
That was a lot of fun. I'm not sure whether they do that now. I~mean there
certainly are discussions that go on. Back then, it seemed mostly in a spirit of
friendliness, but now there seems to be real antagonism in the discussions as
well as in referees' reports. They would make some strong remarks, but I wouldn't
say they were mean then. Being a postdoc in England in the early sixties was
great. We had a wonderful time. During the summer we went to the International
Congress of Mathematicians in Stockholm. I~found that I was reasonably well
prepared for the level of the talks, having been to the various Princeton and
Institute for Advanced Study seminars. It was exciting to see faces \mbox{attached} to many of the names
that I had only read before. Hadamard is one I can mention. I~went to one
lecture in Stockholm---I~think it was Linnik's. I~got there early and talked
with him. After I sat down, in comes Cramer, who sits right next to me!
Then, in comes Kolmogorov and he sits on the other side of me! (both laugh) I was
speechless! As you well know, I~am usually quite talkative. I~guess that I could
have asked for autographs. That would have surprised them I am sure. Sadly I
don't have a photograph to preserve the moment. It was pretty special and
perhaps justified my having gotten a doctorate.

Then, we went back to Princeton. Lorie was pregnant so our life was going to
change a lot. I~went back to a job that was half time at Bell Labs, as Member of
Technical Staff, and half time as a Lecturer in Mathematics at Princeton,
teaching. The two positions were complementary in important ways. Tukey had
created such a structure for himself; however, he was probably half-time in
Princeton, half-time at Bell Labs and half-time in Washington. I~guess that I
then set out to have my own research career. I~had done some writing of papers
before, but now I settled into a more adult research program.

\textbf{Victor:} You seemed to be quite spread out at the time, I~can
see stuff in asymptotics
(Brillinger, \citeyear{brillasym}), Lie group invariance
(Brillinger, \citeyear{brilllie}), fiducial pro\-bability
(Brillinger, \citeyear{brillfiducial}),
resampling
(Brillinger,\break \citeyear{brillresampling})$\ldots.$
Really going off into many directions.

\textbf{David:} Well that was based on material I had learned. I~would
pick up a journal and see somebody had done something and if I thought there
would be a way to contribute, I~would try. The Lie group material was motivated by
Don Fraser. He was creating this area he called structural probability. I~was
trying to see if fiducial probability could be more formalized. R. A. Fisher
kept pushing the idea of fiducial probability. It seemed as if in all his
examples the fiducial probability was a Haar measure. So that was a natural
thing to do. The Lie group paper arose also because people had wondered whether
or not working with the correlation coefficient would lead to a fiducial
distribution. I~showed there was no prior---at least no Lie group measure that
lead to one. But I was still solving problems, minor ones I suppose.

\textbf{Victor:} You mentioned reading papers and thinking about
problems. I~remember reading Tukey's \textit{Statistical Science} interview (Fernholz and
Morgenthaler, \citeyear{stephan}) where he said
that he would pick up journals and read papers, but not really study them. Which did you do?

\textbf{David:} I think I read them over. Because I had a~reasonable
memory and I could read quite quickly. So, a lot of my life has been working on
something and then suddenly thinking, ``Oh, yes, I've seen something like that
before$\ldots.$'' That's a problem with changing universities: because in the Princeton
library, I~might have picked up some journal, but then having moved on to, say,
LSE, I~had to search seriously. Anyway, I~would pick up some journal, and read a
paper that I sought in it, then, just as I was taught to read the dictionary, I'd look
at the paper just before and the paper just after. That way you build up your
knowledge.  Also, when I~have a journal issue in my hand, I~don't think I read
it to study it; rather, I~read it to enjoy it.

\textbf{Victor:} And then came the baby and a decision to make: moving
back to England.\vadjust{\goodbreak}

\textbf{David:} Yes, that's right. Returning was an easy decision.
Because Lorie and I both had loved living in London. Her being from New
York city, and me from Toronto, we were used to, ``Which movie do we want
to see? Then, where is it showing? OK, let's go!'' Princeton was a small town and
Lorie felt pretty restricted. Now we had the baby at home, but her parents lived
up near New York City. I~think it was pretty hard for her. Now women do keep
working, albeit part time or volunteering. But back then, they were right in the
middle of the world, interacting with many people and ideas. Then, all of a sudden,
they were at home for many hours with a baby. Well, Jim Durbin wrote me about
there being a lectureship at the LSE, and was I interested. I~think Lorie and I
just had to look at each other for a moment to know we were interested. I~stayed at Bell
Labs through that summer to finish some projects and to build up
some savings to go to England with. We had a VW van, so we were ahead of the
hippies, and we shipped it over with us. We were driving around London for six
years with this left-hand drive big red VW van.

I have remarked many times that Bell Labs was the best job I had had in my
life. Stimulating facilities, stimulating colleagues, stimulating problems and
minimal restrictions on what one worked on. It is just that Murray Hill was in
the middle of New Jersey. We were very fortunate to have the opportunity to
decide how important was the choice of job as compared with the choice of where
to live. My salary went down considerably of course.

\textbf{Victor:} What was life as a lecturer at the LSE like, and what was
the contrast with Princeton?

\begin{figure*}

\includegraphics{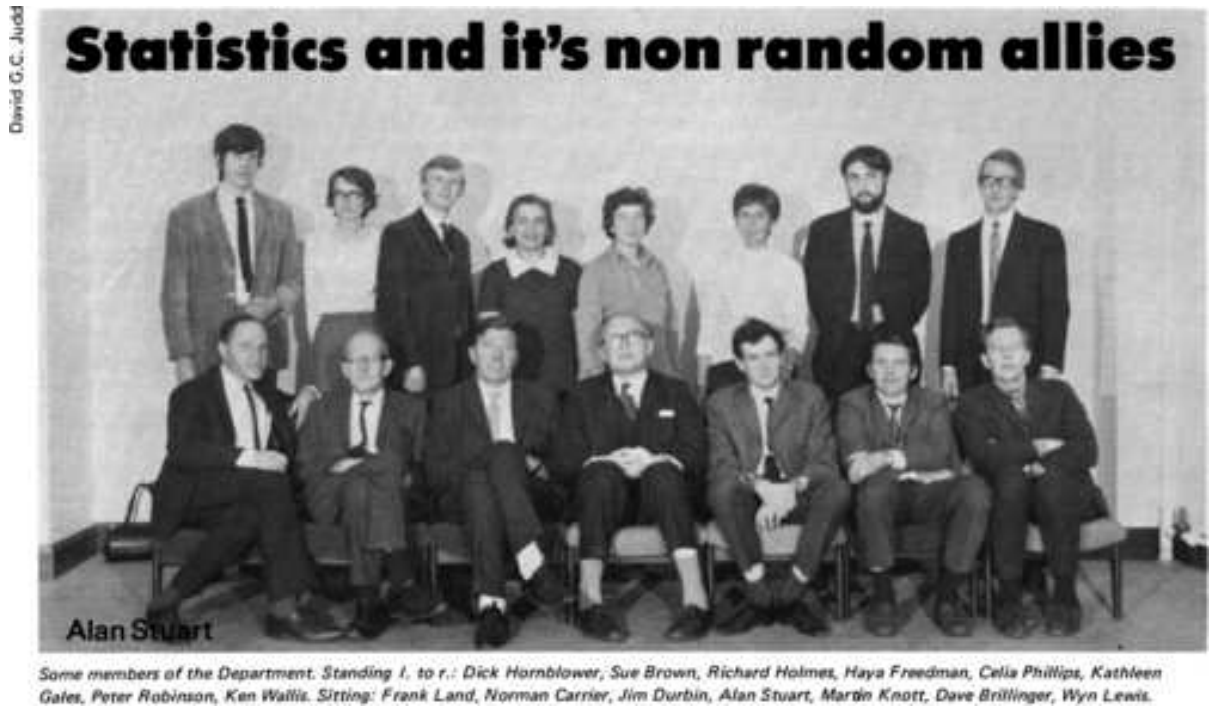}

\caption{The Statistics Department at the London School of Economics in Fall 1969.}\label{lse}
\end{figure*}

\textbf{David:} Well, there were students of both sexes in the
classroom at the LSE! They were left, not rightwing. In both cases, the students
were very bright. Bill Cleveland was in a class that I took over when Sam Wilks
died. Princeton and LSE were very different in many ways. I~did prefer the
English system in important ones. The thing I remember most about LSE is that
there were five, perhaps six of us, who were lecturers at the same time. We were
of about the same age, having kids at the same time, watching the same TV
programs. When Monty Python came along, we would all be talking about it the
following Monday morning. They were teaching me about football/soccer and were
learning about hockey and frisbee from Alastair Scott and me. We pretty much
have all had  successful careers. Fred Smith became the President of the Royal
Statistical Society, Alastair Scott went back to New Zealand and was elected to
the Royal Society of New Zealand, Graham Karlton moved to the Survey Research
Center at the University of Michigan and became prominent in the US survey
community, Wynn Lewis died young, Ken Wallace, the econometrician among us, was
elected a Fellow of the British Academy
(Most of the LSE statistics group in Fall 1969 are pictured and
listed in Figure~\ref{lse}). We were all together, all the time. We
would go to the morning coffee, then have lunch and then afternoon tea again
together. We drove across and around London to visit each other. At Princeton I
was pretty much alone as a young person doing statistics.

\textbf{Victor:} But did your decidedly mathematical outlook tie in
well with what was expected to be published in the British stats journals at the
time?

\textbf{David:} I think that I know what you have in mind with that
question. Just before we moved to England, I~had submitted a paper to the
\textit{Series B} of the \textit{Journal of the Royal Statistical Society}. It wasn't all that
complicated, it was doing factor analysis with time series, getting latent
values of spectral density matrices. I~had in mind the problems Tukey had had me
thinking about, concerning a signal from an earthquake or an explosion coming
across an array of sensors. In an appendix, there was a derivation of approximate
distributions of spectral estimates using prolate spheroidal functions, which
Pollack and Slepian had come up with (Slepian and Pollack, \citeyear{pollackslepian}). The referee said
he didn't understand it and the paper was rejected! And I mean back then I
didn't know about protesting an Editor's or Referee's decision. I~probably
should have rewritten it and sent it back to JRSSB, but what does it matter? I
did give a talk at an RSS meeting. Eventually, I~put it on my website, and it's
still there now. I~developed the dimension reduction aspect further and have
a~paper on that in one of the multivariate analysis symposia and a chapter in my
book. I~don't think this occurrence affected me too much, but some of my
students have been very disappointed by similar things in their career. Best I
can tell them is that parts of life are arbitrary, resubmit.

\textbf{Victor:} By that time, you had been doing quite a~lot of work
on spectral analysis and then in '65~ca\-me the influential paper on polyspectra.
That sounds like a Tukey term.

\textbf{David:} Yes, that is a Tukey term. One of the first things
Alan Stuart said to me in London---you know how picky the English can be---was,
``David, poly is a Greek prefix and spectrum is a Latin word. You are
committing linguistic miscegenation!'' He was just teasing me. But in Volume 1 of
Kendall and Stuart (Kendall and Stuart, \citeyear{kendallstuart}) they say this against Tukey regarding
``$k$-statistics.''

\begin{figure*}

\includegraphics{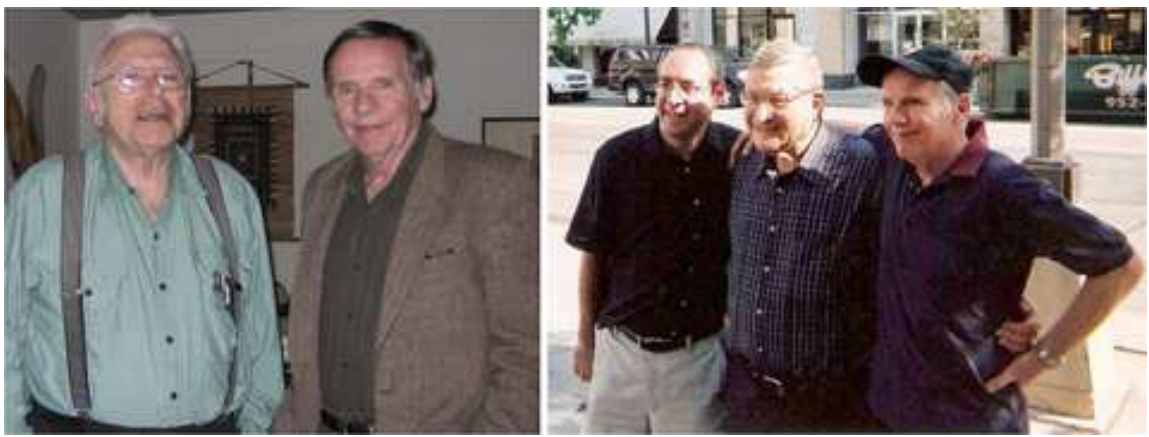}

\caption{David with Murray Rosenblatt, and with Emanuel Parzen and his son, Michael.}\label{fig9}
\end{figure*}

\textbf{Victor:} Surely, there are many such examples---I~can think of
the word \textit{bureaucracy} off the top of my head$\ldots.$

\textbf{David:}
$\ldots$there's another thing that's wrong with
bureaucracy! (both laugh) But anyway, I~mean I was into all this nonlinear
stuff. Tukey, in an early memorandum, had done something on the bispectrum. So
that motivated me to do some research. You know, when you have a math background
you seek to generalize things, to abstract them. It turned out I was unknowingly
at first competing with the Russians---like Sinai and Kolmogorov---when I was
doing that work. I~heard that Kolmogorov had said some nice things about my work
from Igor Zurbenko. That was really nice. Later on, the Russians translated my
book into Russian. I~learned to read Russian mathematics in a fashion, in
particular, the works of Leonov and Shiryaev. That's what got me into the
ergodicity results. For example, what I talked about today was the Chandler
wobble. Arato, Kolmogorov and Sinai had a paper using stochastic differential
equations to explain that motion (Arato, Kolmogorov and Sinai, \citeyear{arato}). I~was strongly influenced by
French mathematics and a lot by Russian probability. I~read the journals of both
regularly. The work on cumulant functions and polyspectra let me get away from
the restrictive assumption of Gaussianity in much of my later research.

\textbf{Victor:} Then, into the picture must have come Murray
Rosenblatt, judging from your three joint papers on higher order spectra
(Brillinger and Rosenblatt, \citeyear{brillrosen1,brillrosen2,brillrosen3}). I~suppose he was in touch with
the Russian school.

\textbf{David:} Oh yes, for sure. I~had met Murray in New Jersey when
he consulted at Bell Labs in 1963. I~remember they had him working on the
cepstrum, which is the inverse Fourier transform of the log of the spectrum.
That work was part of estimating how deep earthquakes and explosions were, and
so on. Then, Murray came to London. And again, I~didn't know I shouldn't do
something like this, being a young jerk, but I just went up to Murray and said
something like, ``How about we write a paper and do some work together?'' And he
said, ``Fine.''  Murray has been my statistical role model, in many senses. Tukey
was a creative role model. But at one point he said, ``Well, David, now that you
are finishing, what do you think you want to do?'' He might have thought that I
still wanted to become an actuary. What just came out of my mouth was, ``I~really don't
want a life like what you have and I am concerned about whether I
want to be an academic.'' And then Tukey put his hands on his chin as he would
often do and said, ``What about Willy Feller? He has a pretty good life.'' So, he
found a role model more to my liking. But then, I~found Murray Rosenblatt. He
just seemed to love his wife and his kids and had a lot going on in his life
outside academia as well as a fine academic career. So he was a good role model. I~don't
think I really managed to express that to him until Richard Davis and I
interviewed him for that article in \textit{Statistical Science}
(Brillinger and Davis, \citeyear{rosenblatt};
see Figure~\ref{fig9}).
He was a lot more of a mathematician than me, but in terms of his life, and
interacting with people, I~respected him.

\textbf{Victor:} Am I right that you also met Emanuel Parzen in
England?

\textbf{David:} Oh yes, and we've been continually in touch since (see Figure~\ref{fig9})! We
also met the Chernoffs then. This year, 2010, Manny and Carol are moving back to
Palo Alto to a retirement home. So we expect to see a lot of them even though
Palo Alto and Stanford have gotten steadily farther apart during our Berkeley
years,
in part because of the growth in traffic. But, with the Parzens moving there, I~expect
Palo Alto to come much closer. Manny and Carol are role models for us in
different ways. One is being a loving couple that were equal, with
each member of the couple helping the other. And the other is Manny certainly
helped me a lot by getting invitations to conferences, and by describing
research that someone else was doing, so I was being kept up. And I think
also by describing my research to other people. He was really the troubadour who
was carrying the information of what was going on in other places around.

\textbf{Victor:} While maintaining a very strong concentration on
cumulants and polymeasures, you also did some things on economics on the side.

\textbf{David:} Bell Labs had a lot of signal processing, so I~was going
into spectral analysis in detail. I~think Kolmogorov and Sinai defined cumulant
spectra in some sense, or cumulant functions. These functions turned out to
provide a natural way to describe ergodicity and asymptotic independence. That's
what I grabbed on to. That was the '65 paper. I~think I might have been the
first one to show that spectral estimates were asymptotically Gaussian without
assuming that the time series itself was Gaussian. The economic work started in
Princeton. Clive Granger---the Nobel prize winner---was at Princeton before I
went to London. He and Michio Hatanaka were working on a book on spectrum
analysis of economic series with John Tukey providing advice. When I moved to
England, Clive was also there, at Nottingham, and would come down to the LSE
every so often, so we had some contact over important periods. Hatanaka and I began
working together and wrote a paper (Brillinger and Hatanaka, \citeyear{brillhatanaka2}). I~presented the
work as an invited talk at the First World Econometric Meeting in Rome in 1965.
Milton Friedman made the invitation. The work was concerned with the permanent
income hypothesis and we had developed a time series spectral analysis
formulation. After the talk, Friedman came up and said something like: ``I didn't
understand any of that but I am sure it was good!'' (laughing) There is another
paper with Michio (Brillinger and Hatanaka, \citeyear{brillhatanaka1}). Data analyses were involved. My
period at the LSE was by far the most theoretical in my career. I~think because
the time series data just weren't there. I~was working as a consultant with the
seismology group at Blackness. It was an offshoot of the Aldermaston Atomic
Weapons Research Establishment outside that base. At one point, I~provided an
effective scheme for them to use with array data, but I guess that I wasn't able
to explain it well enough. That's often been the story of my ideas. I~don't
know, Manny Parzen once quoted someone as saying, ``First you have an idea and
then you go out and sell it.'' But that was never me. I~do try to ask myself,
``Why am I writing this paper?'' In the end, I~think that I am writing for John
Tukey.

\textbf{Victor:} You've often mentioned the influence of scientific
heroes.

\textbf{David:} Feynman would be one. I~have read a lot by
him and about him. I~know that he enjoyed going to Brazil, as I have.

\textbf{Victor:} You didn't have a chance to meet him at Princeton,
though.

\textbf{David:} No, he was long gone. He was there in the early war
years, and left during them for Los Alamos. He ended up at Caltech. When I was
asked to give a talk in Caltech once, he had died before. I~might have been too
intimidated to go talk to him anyway.  Although I~did talk to$\ldots.$  Goodness,
probably you know the name better than me. Who's the MIT linguist, who is in the
news all the time?

\textbf{Victor:} Chomsky?

\textbf{David:} Yes, Chomsky! I took Chomsky out for coffee once. It
turned out that he and Tukey had organized a seminar on linguistics at the
Institute for Advanced Study. This was when I was doing all these memorial
articles about Tukey (Brillinger, \citeyear{brilltukeynotices,brilljwt}). I~had noticed
that Chomsky came to Berkeley regularly. So, I~called a mutual friend and asked
if they could arrange for a meeting next time Chomsky was in Berkeley. They did.
Eventually, I~met Chomsky at the linguistics department and took him
over to this coffee place run by Palestinians. Victor, you have been there. While we
were there, all these people were looking at Chomsky. One woman couldn't resist
expressing her admiration for his work.  He was such a humble, sweet person. I~asked him
whether Tukey had any impact on the seminar. Chomsky said he sat
there and grinned. I~guess one takes that for what it is! So, being a Tukey
student has given me entr\'ee to countless situations. I'll tell you a story
concerning that: just as I was finishing my studies at Princeton, I~was invited to
speak at the University of Michigan---I am sure due to Tukey interacting with
Jimmy Savage. Jimmy Savage did a bit of political analysis of Lorie and me, and
decided that our politics were on the left. He quickly organized for
us to meet with Leslie Kish, sociologist in the Survey Research Center.
That's when our close friendship started.

\textbf{Victor:} Leslie Kish had fought as a volunteer in the Spanish
civil war.

\textbf{David:} That's right, and he was a leader of the Campaign for a
Sane Nuclear Policy. So, Leslie had come to London and was giving a talk
somewhere there. He later told me that he saw that I was in the last row doing
something else. He said he got annoyed, but then immediately thought, ``Oh no,
he is a Tukey student, so that's all right!'' (laughs) Now actually I was
listening! Tukey could do three things at a time, I~could maybe do two,
sometimes.

\textbf{Victor:} Another name you often mentioned is Da\-vid Cox.

\textbf{David:} Oh, yes, he is another hero of mine. He too visited Bell
Labs when I was working there. He was not a professor yet. He clearly had
special things to say. Others might have done some of the things he did in a
more mathematical way and subsequently gotten their names attached to them. I~don't think he
had a problem with that. I~am thinking of things like getting
approximate distributions of maximum likelihood estimators when the model is
incorrect. He did that early on in a Berkeley Symposium paper (Cox, \citeyear{cox}). Then,
in another Berkeley Symposium, Huber came along and did it in a more formal way.
Cox's paper has a wonderful statement, ``Discussion of regularity conditions
will not be attempted.'' There were very few, if any, of David's talks or papers
that didn't have something clever in them. It's as if when he did something, if there
wasn't anything clever in it
(David thrusts his hand as if throwing away a piece of paper), then, no! Out of the window.
He does it all in a very humble way. I~have been on several committees with him and he would say few things for a
while, but he would accumulate information and then he would come up with a
proposition: ``Well you could say $\ldots$ maybe we could do$\ldots.$'' And everybody would
agree. He could merge a lot of different opinions and information. He is one of
my statistical heroes. He did reject a couple of papers that I~submitted to
\textit{Biometrika}. I~took that as saying, you can do better.

\begin{figure*}

\includegraphics{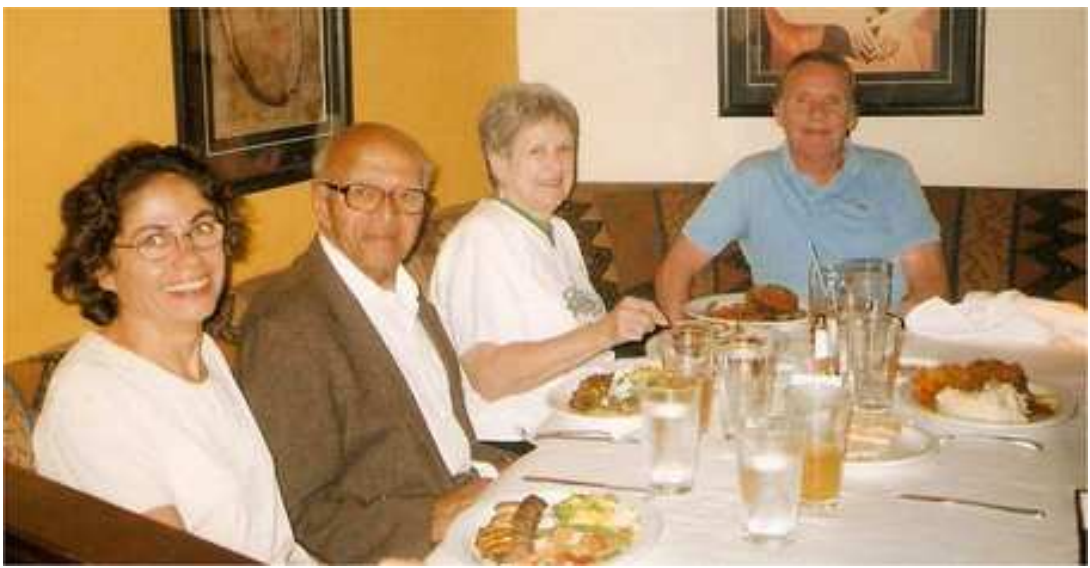}

\caption{David with Lorie along with David Blackwell and Maria Eulalia
Vares.}\label{fig10}
\end{figure*}

\vspace*{3pt}\section{Going to California}\vspace*{3pt}

\textbf{Victor:} I understand that you would have been~ve\-ry happy to
stay in London, but then things changed.

\textbf{David:} Yes, well my mother retired. She had had a~hard
life. She was a very bright woman, but because my maternal grandfather died in
the great flu epidemic leaving my grandmother with five children, my mother had
to go to typing school to help the family survive. Many years later, she went to
adult school and got to be a country schoolteacher. We were sending her some
money, but when she retired her pension was tiny. Even though I had become a
Reader at LSE, there was just no way I made enough to make up what she needed.
We had Jef and Matthew at that point, we were living quite happily, had a nice
house a block away from Wimbledon Common. We were going to the theatre and
concerts regularly. But there just was no way to be able to also support my
mother. So I had to look for a higher income. Berkeley had already invited me
several times. Actually, David Blackwell had called me just before I finished at
Princeton (see also Figure~\ref{fig10}). Now in the late sixties Berkeley was the place
to be, with the free
speech movement, rock concerts, experimentation in the arts and all that. We had
learned that when we were there on sabbatical in '67--'68. There were a growing number
of protests against the Vietnam war, and Lorie was quite involved. So we knew
Berkeley, and they knew me. And when Henry Scheff\'e asked me about moving there, we
agreed. A person high in the academic totem pole told me once that a senior
department member had said that I was the most influential appointment in the
`70s. There were lots of mathematical things going on and I enjoyed that, but I
was strongly interested in applications of mathematics. I~immediately fell into
place with Lucien Le Cam and Jerzy Neyman and all their visitors---they had a
lot of important ones. So, we left London because we needed a higher income, but
we landed in a very special place. Our older son, Jef, loved England. He was
very sad about the move and that made Lorie and me sad. I~think we expected that
eventually he would move there.

\textbf{Victor:} So tell us a bit about your early Berkeley years.

\textbf{David:} The earliest years were '67--'68 when I was a visitor on
leave from LSE and we have already talked about them. We moved to Berkeley
permanently, arriving by ship, in January 1970 to be met by Erich Lehmann on one
of the piers. At that time, there were a number of individuals who were then
Assistant Professors but who did not get promoted to tenure, that is, had to pack
their bags and leave town. They were able academics so their nonretention was quite a shock
for me. Actually, it seemed inhumane. Some of these
people had children already at school. I~was used to the English system where,
if you were a Lecturer, and you had passed across the bar after three years,
then you had tenure. You would hit the top salary of the lecturer scale but you
might stay in your department the rest of your career---you had tenure. Some
people did take advantage of that. We lost Berkeley friends that we had made and
that was a great shock. Apart from that, we were really enjoying the department,
Berkeley and the Bay Area. The department seminars and the quality of
the discussions in the lunch room were top notch. In these early years Kjell
Doksum and his family became close friends.

\textbf{Victor:} Did you thus quickly forget about London?

\textbf{David:} No, not really. In fact, when in 1971 David Cox wrote
that a professorial chair was available at Imperial College, and asked if I was
interested, I~was very interested! But going through the sums, with Alan
Stuart's help, we just could not afford to return. Our old house was now worth
more than twice as much as we had sold it for, within that short period. We
couldn't afford to buy a comparable house.

I have sometimes wondered how things would have worked out with Jef's brain
tumor had we returned. Cormack had just developed the first CT scanner at
Atkinson Morley Hospital just down the hill from our Wimbledon house. That
technology wasn't yet available in the US, and might have helped.

\textbf{Victor:} But you found data at Berkeley.

\textbf{David:} Yes, I~found data and fine applied scientists to work
with at Berkeley. On reflection, I~had reached the career that Tukey and Bell
Labs had been training me for. Soon after arrival, I~just wandered over to the
seismographic station where I met this Australian fellow, Bruce Bolt. He and his
family became dear friends. He was a sailor also, so we spent time on the Bay in
his boat. Our families mingled. Bruce was religious, and I was no longer.
However, we didn't seem to have the slightest difficulty talking about religion
and other serious topics. He got me working on time series and other problems in
seismology. We wrote several joint papers, but affected each other's research
quite generally.

\textbf{Victor:} Was that around the time you wrote your invited paper
on point process identification (Brillin\-ger, \citeyear{brillppsystem})?

\textbf{David:} There is a history to my work on point processes both
in London and Berkeley. David Vere-Jones, another dear friend, another
influence, presented an Invited Paper at a meeting of the Royal Statistical
Society (Vere-Jones, \citeyear{VJ}). I~was asked to second the vote of thanks. When you are the
seconder you are supposed to criticize the paper's content. Victor, you've
probably been to these things. So I read David's very seriously. I~don't think
I had much in the way of criticizing, but it got me very interested in temporal
point processes.

At Berkeley, Neyman and Scott had done path breaking work on spatial point
processes, particularly in astronomy. Six months after my arrival in Berkeley in
January, the Sixth Berkeley Symposium took place. I~presented\vadjust{\goodbreak} a paper showing a
way forward for making inferences based on data for processes with stationary
increments (Brillinger, \citeyear{brillberk}). This included stationary point processes. Around
that time I also had a student, Tore Schweder, who was looking into that point
process material when modeling whale tracks. To continue the story, while Betty
Scott was still department chair she asked me if there was anyone it would be
good to invite to Berkeley for a term. I~suggested David Vere-Jones. He and
Daryl Daly came, and a whole world of point process work got started. In
particular, David and Daryl organized a seminar series. Peter Lewis and ``Pepe''
Jos\'e Segundo were important speakers. Peter's energy and enthusiasm and broad
knowledge captivated the audience. Pepe came with specific problems and data
concerning the firing of nerve cells. Pepe was a Professor in the Brain Research
Institute at UCLA. And he had all these wonderful data on nerve cells firing.
And I just said, well this model that I have been fitting for earthquakes might
be good. So then he sent me these massive piles of boxes of computer cards! They
took up perhaps 10\% of my office for many years! The thing that was interesting was
that second-order spectral analysis seemed to be quite effective. So, I~was
working on point process data from seismology and point process data from
neurophysiology at the same time. My students Rice and Akisik worked on these
models/data also. The advantage of the neurophysiology case was that it was a
designed experiment situation and, thus, you could repeat the experiment. So, that
collaboration resulted because I was working on point processes from seismology.
To my mind, one of the major successes was that the concept of partial coherency
analysis could be extended quite directly to the point process case
(Brillinger, \citeyear{brillppsystem}), and it let one infer the causal structure of networks of
neurons (Brillinger, Bryant and Segundo, \citeyear{brillbiol}).

Pepe had a daughter who died in a plane crash at Puerto Vallarta. At that time, I~had a son with
a brain tumour that could not be removed. These tragedies brought
us very close together. Having\break a~child die is pretty hard. Pepe and I had our
scientific conversations to keep us focused on one good side of life.

\textbf{Victor:} Would you like to talk about Jef?

\textbf{David:} (David pauses and speaks with a broken voice.) Well,
yes. I~mean it really affected Lorie, Matthew and me, as well as Jef's and our
friends. We have cared a lot about other people always. I~don't believe\vadjust{\goodbreak} that it
is an accident that Lorie became a~nurse midwife or that I started working with
nerve cell spike trains. One works to fight for political ideals and to improve
the system, but it is totally humbling to care so much about a child and not be
able to help them in their time of greatest need.

Jef's illness went on many years. The first hint was in 1968 and he eventually
died in 1988. It was not diagnosed as a brain tumor until 1973. He had
three bouts of brain surgery and radiation between 1973 and 1988. In 1973 he was
supposed to die within 6 months, but he just kept coming back. The night he died, I~didn't think
he was going to die. He graduated from UC Santa Cruz in 1988,
just two years behind his class. Everyone did everything imaginable. The
doctors, his brother Matthew, Lorie and her nursing friends, our friends. The
doctors made home visits. Nobody wants to see a child die. Many, many people
attended the memorial.

Jef had a motorcycle, just as my mother and father had. I~sometimes think about
his motorcycle. I~knew that I wasn't going to get on it, but I knew about it. Jef
rode it back and forth to Santa Cruz, in part over a mountain. Once, there was a
heavy rain storm and he thought that he might die. Another time, someone in the
back of a pickup truck threw a bottle at him. He could have died on that
motorcycle so easily. Then it would have been: if only, if only, if only$\ldots.$
That's what our memories would have been. But our memory is that everybody did
the best they could. Including Jef. Lorie has been really hard hit with death.
She's had to nurse her dying parents, her son and her sister now.

\textbf{Victor:} Practically everybody who's met you will attest to what an
uplifting person you are, how it seems that you are always smiling.

\textbf{David:} Not always but most of the time. Probably my life was
all fun until 1973 when Jef was diagnosed with the brain tumor. Science and
researching kept me going through those times. Nowadays, I~just have to think
about my grandchildren and a smile surely appears on my face. Having gone
through all this, I~do go to a lot of effort to communicate with the Berkeley
students about the importance of enjoying every day and realizing how lucky they
are. In one of my classes in Berkeley, I~realized that I was assigning a great
number of problems. What I did at the spur of the moment was to say, ``OK, your
problem assignment for this week is to go to a movie and then write on a piece
of paper the name of the movie you've been to!'' I think they just thought I was\vadjust{\goodbreak}
kidding. I~wasn't. I~have a hard time convincing today's students to put things
into perspective. They seem quite terrified and not having all the fun that I
had as a student. They are overly worried about getting registered in a class,
about finding a~thesis topic, about getting a post-doc, about getting a job,
then about getting tenure, about getting a grant, getting to be a professor,
getting to be invited to conferences. They have the problems of old people on
their shoulders already! I am just sad for them. Things do work out. I~hope
you're trying to get your students to enjoy life, follow sports, things like
that!

\textbf{Victor:} Well,  I've had good advice, and try to pass on
what I learned. Did research and sport help you at all during that difficult
period?

\textbf{David:} When I was recently preparing an encyclopedia article
on ``soccer/world football''---that was the title I was given---and I was pulling out
a lot of books, I~found that there was a book by a couple of Russians on
applications of mathematics to sports (Sadovski\u{\i} and Sadovski\u{\i}, \citeyear{russianbook}),
because it has some material
on soccer. When I read the introduction, I~found them saying that to do
mathematics well, you want to be healthy and fit. I~have known this for
many years, but it was reassuring to see it in print. I~think that participating
in sports is important. You know, running around and interacting with
others. I~think of Shiryaev, since we're talking about the Russian point of
view. He is a~very good skier. He received a medal for it. There is something
specific I'd like to feed into our conversation just now. I~played a lot of
intramural and informal soccer over the years. One year, two teams the Statistics
Department was involved with met in the final. However, I~stopped playing after
Jef died. I~wanted to be alone. Friends would come by my office to try to get me
to play, but I just wanted to be alone. But my office looks over the Bay and
much of the time I could see people sailing and windsurfing. I~thought, ``Why
don't I try windsurfing again?'' I had tried once before and it hadn't really
stuck. But when I tried again, I~got the basics. Windsurfing is one of those
things where if you don't know what to try to do, then you are in big trouble.
What I found personally was that if I thought of anything else when I was
windsurfing, I~would fall into the water. After I windsurfed for 2~hours I was
just high. One day when I went back to Evans Hall, I~saw Andrew Gelman and said
something like, ``I windsurfed all the way to Emeryville today!'' Andrew said,
``Well, I~climbed up the outside of Evans Hall today!'' (laughs) It was that male
thing, if someone is boasting too much, they get brought down. I~do recommend to
anyone who has some tragic situation to deal with, and they do like outdoor
activity, that they take up windsurfing.

\textbf{Victor:} What was it like to arrive at Berkeley in the late
60's--early 70's?

\textbf{David:} Super. Rock concerts, progressive politics, long hair,
hippies, tear gas. I~was teaching once in a room in Wheeler Hall and all of a
sudden there was some strange unfamiliar smell. I~didn't know what was going on
until someone in the class said, ``That's tear gas!'' It was really something. There
had been ``troubles'' at LSE, but none with tear gas. I~remember one friend I
have, especially. When there was something radical going on I was out of there,
headed away from the trouble. But I would invariably see him heading the
opposite way, that is, in the direction of the trouble. I~did see some bad things.
Through my then office window on the third floor in the Physics building, I~saw a
sheriff's deputy club a young man who was just sitting under a tree reading a
book. I~think officers were totally frustrated because the demonstrators where
leading them in a chase across campus. I~do have to say that some were throwing
rocks---and that's not cool. The deputies chased but they could not catch these
guys. So, they just got more and more frustrated. Here's another story from that
time period. Al Bowker had become Chancellor and joined our department. He had
to deal with various ticklish situations during his tenure. Somehow, he always
found a way. Evans was a new building and its inside walls were stark. One
weekend some of the mathematicians came in and painted some murals. There was
one of the death of Galois. The custodians cleaned them off. But the
mathematicians repainted the murals. A battle of wills was developing. Bowker
said just leave them. Long after the murals were painted over when the building
was refurbished and I don't know that there was any fuss.

\textbf{Victor:} Al (Bowker) told me a story about some students who
were demonstrating. They came into his office wearing dark sunglasses---I
suppose it was some sort of statement. But then Al caught them off guard: to
their surprise, he was already wearing dark sunglasses himself (both laugh)!

\textbf{David:} I had some fun like that too. When I was department
chair, Lorie's brother was working for a video company that had produced a movie
titled ``Take This Job and Shove It.'' He mentioned that they were giving away
hats with the movie title embossed. I~asked if he could get me one of those. He
did. One crisis that developed in my chairmanship occurred when the campus
wished half of our space back---I confess that Betty Scott had been too
effective in getting us space in the new Evans Hall. Anyway, when I went to see
the Vice Chancellor I wore the hat and then passed it on to him! (both laugh) We
ended up losing a quarter of our space.

\textbf{Victor:} What about departmental life? For example, Jerzy
Neyman?

\begin{figure}

\includegraphics{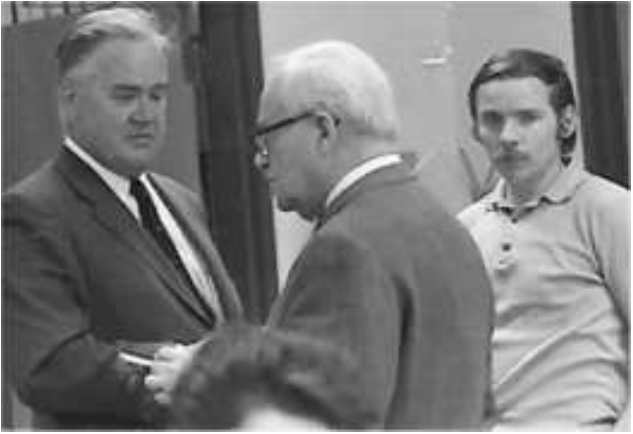}

\caption{David with John Tukey (left) and Jerzy Neyman
(center).}\label{fig11}
\vspace*{-3pt}
\end{figure}

\textbf{David:} As far as I was concerned, being around him was a treat.
One of Neyman's goals was ``to find a model describing the data.'' In contrast,
Tukey's goal was to ``discover surprises in the data.'' Neyman was more for
formalization, whereas Tukey was more for intuition. Surely, both are needed. I~saw the two masters of these things at
work (Figure~\ref{fig11}). I~attended the Neyman Seminar
regularly and went for drinks afterward. Neyman had a host of really wonderful
visitors coming to Berkeley. I~had total respect for that man.

\textbf{Victor:} And Neyman was one of the people you had gotten closer
with, along with Le Cam and Scott?

\textbf{David:} Yes. For one thing, they were always in the coffee room
at lunch time, often with famous visitors eating Neyman's hard boiled eggs. The
talk was lively, what with Neyman knowing so much about European history, all
his languages and poems, and Betty being so full of heart and caring for people;
Lucien being very French in such positive ways. The three cared so much about
the students. Surely, the best part of Berkeley has always been the students.
Once when I was in the coffee room, with Neyman and Le Cam, a student came in
whose father was having a medical problem. Lucien and I were chipping
in suggestions. After listening a while, Neyman remarked, ``Isn't it wonderful
that the professors are helping out the students with their personal problems?''
All three would jump to help with student's personal difficulties. They were
wonderful. I~have been a bit unsatisfied with the Neyman biographies. They don't
seem to bring out the essence of the man. I~said this to Betty and Lucien once
and they agreed. Biographies of scientists, by their nature, seem to focus on the
science side. Setting down the human side is surely much harder.

I'll tell you one of the funny things that came to my head just now: somebody
asked me once if I thought that Betty Scott and Jerzy Neyman were lovers. My
immediate response was, ``I hope so!''

\textbf{Victor:} You had been exposed to two of three main schools of
thought in statistics: Tukey-esque, British and then came the third: Berkeley.
What was that encounter like?

\textbf{David:} I would like to start by replacing ``Tukey-esque'' with
Tukey-Bell-Labs-esque.  That's the\break school that I learned EDA in. OK the
encounter. I~start by quoting Le Cam at this point. Once, at lunch, I~told
him about some research that I had just seen suggesting that cigarette smoking
wasn't bad for one's health and at about the same time another report that
suggested it was bad. What did he think about that? He replied, ``They're both
right!'' The three schools are all right. We need each. I~think it is important
for people to travel and experience all three. The RSS meetings, for example,
are a way to learn the British school. One meets these people and compares their
discussions of the same paper. A lot of things exist in the scientific air, but
are not written down, particularly heuristics. And it's very important to have
heuristics along the way to nailing a~problem down. Often, when you go to another
center and are in a discussion, they quickly draw a little diagram and then you
have picked that representation up. The thing is that you could go a whole
career and never know that something could be simplified that much. As the years
have passed, the British statistics school has become a lot more American. For
example, consider measure theory and theorems.  There have always been a lot of
wonderful probabilists in England, but they did not appear to have much influence
on the statisticians until recently. One thing that I particularly respect about
the English system, including people who aren't famous, is how well they can ask
questions. There would be someone at a seminar, and then there would often be
someone with a British accent who would put their finger on a crucial point
that's going on in the science. Not so much the mathematics, but the science of
the situation. I~have a lot of respect for that. What was the encounter like? I
flitted among each of these schools. I~am a scavenger. I~have the luxury
of trying a Tukey approach, trying a Cox approach and trying a Neyman
approach to problems. The Bell Labs group was influenced strongly by Cox, by
Kempthorne and by Tukey. They weren't much influenced by Berkeley or Box.

\textbf{Victor:} 1975, \textit{Time Series: Data Analysis and Theory}
(Brillinger, \citeyear{brillbook}).

\textbf{David:} Well, that book has got blood on every page! I wrote it
when I was in England during the late sixties. It took too long to be published. I~did
enjoy working on it. I~was going to LSE two days a~week. We had a three-story townhouse.
I~would sit down on the top floor listening to the BBC's
wonderful radio programs, working away on the book, while Lorie would be two
floors down with Jef and Matthew. In the afternoon, I~would be all involved with
the kids. It was so enjoyable. The book started from my research, which got
simplified for my lectures at LSE. Before reaching Berkeley in my 67--68
sabbatical, we spent the summer in Princeton.
Tukey and I were supposed to be writing something up. But Tukey decided to go
off somewhere, and there I was at Bell Labs. Ram Gnanadesikan asked me to give a
course on time series. Luckily for me, somebody at the Labs was available to type
up the notes. This provided a fine start to the book. There were all these
wonderful computing facilities. The fast Fourier transform, a fast computer and
graphics all came together there. Then I got back to England in the summer of
1968 and I guess that's when the serious filling in of material was done. The
manuscript went to the publisher in '72 after I had made a serious attempt to
have the references complete. It was printed in '74, but they put a
date of '75 on it. It has now been with 4 publishers! That sounds amazing, but
Holt--Reinhart gave up their statistics list, Holden-Day went broke, and then it
went to McGraw-Hill who put their binding on it but didn't do
much else. It is now with SIAM and called a classic. How about that?  There were
some surprising benefits, like not having to do much preparation for lectures
for many years. The thing that I enjoyed the very most was making up the
problems at the ends of the chapters. Because I'd\vadjust{\goodbreak} be thinking, ``Maybe there is a
problem sort of like this,'' or ``Maybe reasonable assumptions are something like
these,'' and last ``Maybe a solution could go as follows.'' The thing is one is
negotiating with these three different vague items. It turned out that solving a
problem was a lot easier than creating one! Victor, I~did a vain thing the other day. I~typed
``Time Series: Data Analysis and Theory'' into Google. It claimed to have located
136,000 results!

\textbf{Victor:} You must have taught the time series graduate course
``Stat 248'' at Berkeley for many years.

\textbf{David:} I think every single year, except when I was on
sabbatical. I~believe Bob Shumway came then.

\textbf{Victor:} So did you change it quite a bit? I remember sitting
in on three different versions.

\textbf{David:} Oh yes. I~design it totally differently every year---and no one seems to
notice! To allow variable content, I~call it ``Random
processes: data analysis and theory.'' A couple of students, not you of course,
have said they should have come back. I~try to tie it in to something I'm
excited about at the time. Perhaps trajectories, perhaps point processes,
perhaps spatial-temporal data and so on. I~think if you are not excited about
something, or if it is something you have done a long time ago, it's boring.
Nowadays, there are all these wonderful data sets and graphical devices to
employ. It can take some time to prepare a display, but it would be a great
shame not to.

\textbf{Victor:} You spent some time as a Visiting Professor of
Mathematics in New Zealand. I~know you are in love with New Zealand, is that
when it started?

\textbf{David:} Yes. Alastair and Margaret Scott became dear friends in
London. Alastair and I were Lecturers together. We had met at Bell Labs, and
when I arrived in London he wrote me wondering if there were any jobs. So, I~asked Jim
Durbin, and there was a Lecturer position. Alastair stayed a couple of
years longer than me. When Jef had the first surgery, he was really set back a
long way. We wanted to go somewhere gentle, and that was New Zealand. There, his
energy came back and he could do things like play basketball at a boys club
Friday evenings and come home alone on the bus. He was about 12--13 years old
then. It was the way things had been for me when I was that age. The Scott's
friends became our friends right from the start. Alastair and I tried to
collaborate on a paper once, but we never seemed to talk statistics. It wasn't
that we didn't want to or couldn't, we just seemed to get talking about other
things. But I do believe that we have influenced each other statistically a lot.
So, New Zealand became our home away from home. NZ is where Lorie and I~retreated
to in 1988. That year was horrible. Lorie's father died, Jef died and my mother
died. It has been important to Matthew, too. When Matthew decided he wanted to
do a doctoral thesis in literature on Nabokov, it turned out that the world's
expert on Nabokov was in Auckland! To tie the knot even tighter, we have three
Kiwi grandchildren.

Another place I have a strong connection with is Brazil. It began in the
context of graduate students. I~had three Brazilian graduate students pretty
early in my career. For many years, they were inviting me to come visit. I~would tell
them I was not going to any dictatorship. But, eventually, the generals
went away and luckily I was asked again. I~went that time and had a wonderful
visit. Brazilians and Canadians are very similar in many ways it turned out. In
particular, they both have very high levels of teaching and research in
statistics and, of course, sports are very important in both countries. Then, I~got
invited to another meeting and Pedro Morettin proposed that we apply for a
joint NSF-CNPq (stet) grant. When the grant was funded for 3--4 years I decided
it would be rude to have that grant and not make some attempt to learn
Portuguese and took two courses. I~have given talks in Portuguese there and they
have been very patient with me. One of the days that I was most proud of
professionally was when I got elected to the Brazilian Academy of Sciences. That
was quite a surprise!

\textbf{Victor:} You also chaired the department at Berkeley for a
couple of years. How was that?

\textbf{David:} I liked some parts of it, a lot. I~got to
know the staff very well, which I hadn't before. I~got to know all the grad
students very well, and many undergrads. I~had many pleasant interactions with
my colleagues also. But I couldn't do any research. Because whenever I tried to
do research, all of a sudden the day became too short or I was interrupted too
often. I~had agreed to do it for one year. The ``candidates'' had come down to
David Freedman and me. David Blackwell said, ``Well, it's you two. Time to
choose.''  David and I each agreed to take it on for one year. I~thought it was
unfair that I was being expected to take it on then, because I had so many
projects in process. David Freedman probably felt the same concerning himself.
In the end, I~did it for two years. David F. did it for five. As I just said, I~did
enjoy the job, but only after accepting not doing much research. The person\vadjust{\goodbreak}
whose model I followed in the job was Erich Lehmann.  He had been chairman
perhaps for four years and I just liked the way he did it. He would be in the
coffee room at 10 a.m. in case any of the students or faculty wanted to see him.
One needs role models for how to do these different things, and Erich was my
model for the chair position.

I just remembered a story. Actually, during Erich's term I was (Acting) Chair for
half a day. Erich had felt compelled to resign over some matter. I~was Vice
Chair which I guess made me Chair in a sense. However, Erich didn't tell me that
he had resigned until my ``term'' was virtually up.

\begin{figure*}

\includegraphics{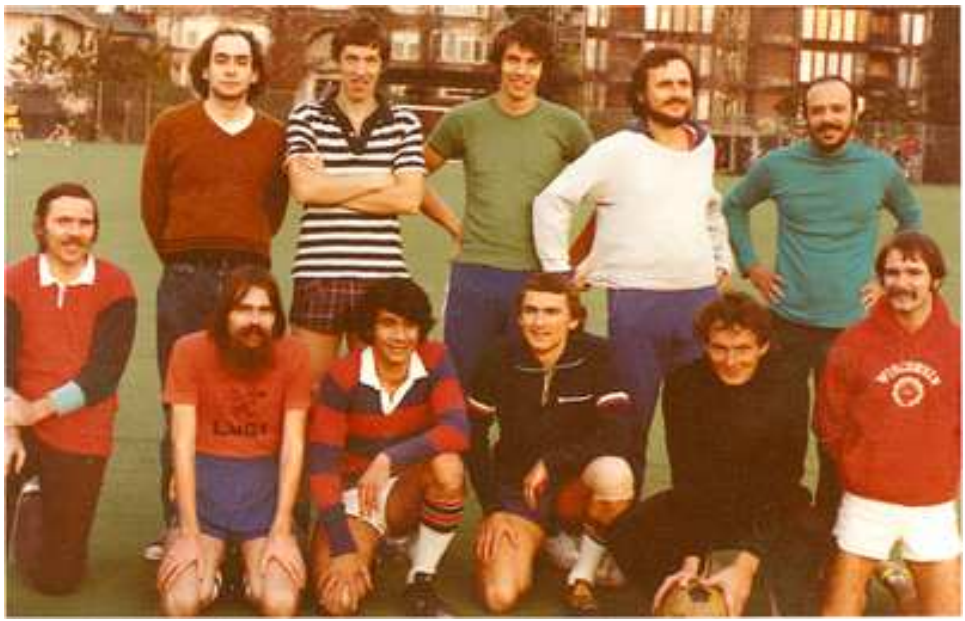}

\caption{David with a group of his Ph.D. students in Banff, 2003. From
left to right, starting at the top: Bruce Smith, Peter Guttorp, Tony Thrall,
Knut Aase, Mark Rizardi, Rick Schoenberg, Ed Ionides, Isuo Miyaoka, Haiganoush
Preisler, Jostein Lillestol, Tore Schweder, John Rice, Andrey Feuerverger, Alan
Izenman, Raju Bhansali, David.}\vspace*{5pt}
  \label{brill_students}
\end{figure*}

\textbf{Victor:} So what is your opinion on leadership in academic
departments? There's a sort of patriarchal paradigm with a dominant personality
at the top and a democratic paradigm---for example, Neyman years vs. post-Neyman years. What's your take on that?

\textbf{David:} There is also an anarchist model. In fact, when I first
came to the Department there was something of an anarchist attitude---everything
was being challenged, like language requirements. Barankin gave a stirring
speech, which got rid of them. I~believe that Neyman created some things that
might never have existed without him. That was very special and what the right
great leaders do. I~don't feel that the faculty resented it too much, but I
don't know. I~liked being at the LSE rather than some other English university,
because then there were something like 5 professors in the department (Figure~\ref{lse}).
Also, mathematics was growing out of statistics there, not the other
way around. The professors rotated the\vadjust{\goodbreak}
position around being chair for three
years. What I tend to say when people tell me that they have been asked to be
chair is: well, if you can do it, you have to. The thing is if the people who
could do it manage to get out of doing so, then the system of good governance
collapses. Anyone who could do it has to take their turn. An advantage is that
different things are emphasized depending on who is the chair. In my term, I~put
a lot of department resources into computing. It seemed the time for that and I
could handle the decisions. Incidentally, one of my students said that as soon as
he learned I was going to be chair, he worked very hard to get his thesis
finished. So my taking the job on was good for him.

There are different attitudes concerning how to behave as chair. When
I was doing it, the budgeting was actually very loose, but I didn't
know that. A~friend who was chair of another department heard me muttering
about restrictions on money. And he said, ``Oh just spend it! Let the dean
find the money!'' I guess there was no mechanism at the time to pick up on
overspending. When I told the financial dean that I was spending money like it
was my own he said, ``Good!'' Many university things were much more casual back
then.

\textbf{Victor:} By next year, you will have had 40 students, some very
notable people among them.

\begin{figure*}

\includegraphics{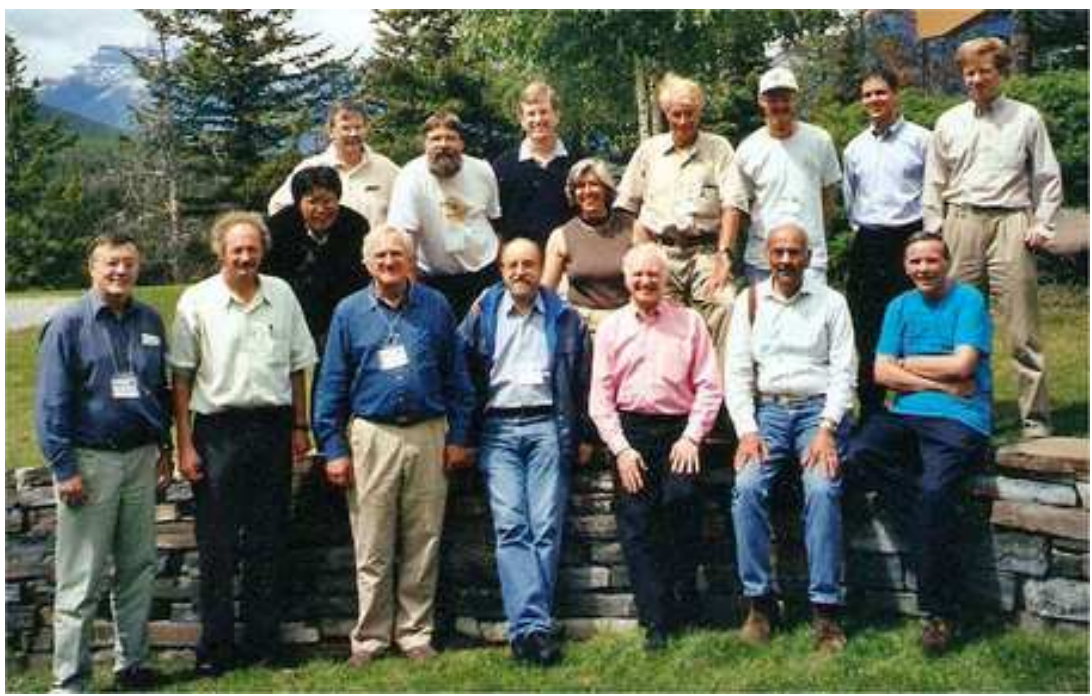}

\caption{David and the Berkeley Statistics Soccer Crew. From left to right,
starting at the top: Tom Permutt, Jan Bjornstad, Jim Veetch, ?, Annibal
Parracho, David, Peter Guttorp, Kai(-squared), Eldar Straum, Albrecht Erle, Ken
Suttrick.}\label{fig13}
\end{figure*}

\textbf{David:} Students have been one of my great joys at Berkeley. If
for no other reason, they are a\vadjust{\goodbreak}
motivation for seeking a position here. There is a nice picture of me with
many of ``my'' doctoral ones in Banff (Figure~\ref{brill_students}). I~sometimes wonder
whether I could have supervised a student and not become
friends with them. They certainly do become friends. As you point out, my
rate is about one student a year, and that's probably a reasonable one because they
take 2--3 years to complete the thesis. Nowadays, there are research groups or
labs. I~tried that in the mid-seventies, but it didn't seem to work well for me,
or, more importantly, for the students. My goal is to have the students learn how
to do independent research. This was Tukey's way. I~sometimes see my ex-students
treating their students the same way. I~interact with a student to find a topic
that they are really interested in. Nowadays, statistics is everywhere, so that
hasn't been too hard. I~think when you are interested in something, you just find
yourself progressing and the time flying by. I~used to play a lot of intramural
soccer (see Figure~\ref{fig13}). That's actually a good way to get to know students and visitors. When
you kick them, accidentally of course, you see how they respond and when they
kick you, they see how you respond. You learn a lot about each other!

By the way, I~will not sign off on a student's thesis until they have started
arguing with me and are calling me David. For some students that can be hard,
but they need to be toughened for the outside world.

\section{\texorpdfstring{``$2\pi\neq 1$''}{``2 pi /= 1''}}

\textbf{Victor:} I was wondering if we could go back to research a bit.
The title you used for your 2005 Neyman Lecture (Brillinger, \citeyear{brillneyman}) was
``Dynamic Indeterminism in Science.'' Would you say this describes your
scientific vita?

\textbf{David:} I like your question. In a word, the answer is
maybe. That expression is to be found in a 1960 paper of Neyman's
(Neyman, \citeyear{neymanindeterminism}). He was encouraging people to learn about
stochastic processes. I~don't think many statisticians did back then. And then
I~was invited to give a talk (Brillinger, \citeyear{brillwarsaw}) at the International Congress
of Mathematicians\vadjust{\goodbreak} in Poland in 1983. I~talked about statistical inference for
stochastic processes in a general way. There we\-ren't many people doing that
then. Murray Rosenblatt and Ulf Grenander were involved with it, but the list of
people working with a general process framework was short. One conceives a datum
that is a realization of a process. That's what Neyman was encouraging people to
work with. Le~Cam's approach was totally abstract, so everything was a particular
case---but in a sensible way.

\textbf{Victor:} I recall you were mentioning in the doctoral course on
applied statistics at Berkeley that, ``Any mathematical object
that can be mathematically expressed is potentially data.

\textbf{David:} For sure. You just put a collection of the objects in a
hat. Then you find a sensible way to pick one of them at random and then you've
got a~realization of a random object. Think about the article I showed at my talk this morning
about statisticians being the sexy thing to be for the next 10 years
(Lohr, \citeyear{nytimes}). The rest of the world has clued into that, finally! There are
these wonderful data sets with people who care about them. And statistics has an
immense amount to contribute to their study. Plus, it's going to be a lot of fun
to be doing it. You have music in your computer, videos in your computer, you may
even have a Bible in your computer---all this stuff is nowadays in a computer,
just waiting for you to discover surprises in it! That's a Tukey attitude. I~never saw
Tukey doing any computer programming, but he could surely visualize
it. And he was very much involved in the first Von Neumann computer
(Brillinger and Tukey, \citeyear{brilltukey}). So, he knew about it in that sense. I~did see him with
coding sheets, but he was preparing things for cards to be punched for his
citation indices (Brillinger and Tukey, \citeyear{brilltukey}).

\textbf{Victor:} Some consider you as a theoretical statistician,
others consider you as an applied statistician. Which one is it? Always learn
new theory?

\textbf{David:} Oh yes? Where did you get that?!? (both laugh)
That's my motto: always learn theory, for the theory becomes the practice. I~can
provide a lot of evidence about that and I think it is what places the Berkeley
students in a good position when they finish. Because other places will create
students who are really up to date the moment they finish, but not ready for new
things that come along. It's harder for them to keep on top of things.
They may well feel intimidated and struggling to keep up. I~think the students
coming to Berkeley get a lot of gifts from the people here. One can mention
Le~Cam with his abstract approach to things and depth of thought. I~had great
respect for him for a lot of reasons. One of them is he could sit in his office
and he could dream of these incredible mathematical problems, and dream up
solutions. Whereas my thing to do is to find a parallel scientific situation
where that problem exists. This can give important clues about how to approach
the problem. Lucien always seemed able to generalize these things in such a way
that he would encompass so many things. I~would take some of his work and
particularize it to a specific situation.

\textbf{Victor:} Is that your research strategy? How do you attack
problems? How do you find or choose them?

\textbf{David:} I find them by people interacting with me, or by my
asking them. As I mentioned earlier, when I~arrived in Berkeley, I~went over to
the Seismographic Station. They didn't come to me. I~think that with a
consulting service you don't really get the special people coming. You have to
go over to them, to the scientists. You have to present yourself to them. Terry
Speed and I agreed on this once. Terry was chasing across campus some time after
he arrived, interacting with people, particularly in biology. When I think about
my recent work: risk analysis was motivated by interactions with Bruce Bolt of
the Seismographic Stations, the trajectory modeling was based on data collected
by Brent Stewart of Hubbs Sea World, while both topics involved Alan Ager and
Haiganoush Preisler of the US Forest Service. The work on sports statistics is
based on data that I collected on my own. At a certain point you've got all the
problems you can handle. It seems in any case that if you want to work with good
people, then you have to go after them. So I've just come to know a lot of
people. Various of my papers may be found in (Guttorp, \citeyear{guttorp}).

Now, I~am a member of the scientific of advisory panel this new center of excellence
for evolutionary biology at the University of Oslo, and there is a flood of new
problems coming into my head from that. It is just wonderful. But I was
wondering: why me on this panel? And then I thought, ``Oh, evolution, that is time-series,
isn't it?'' It is just a totally different group of scientists from any I
have been involved with before. Now I own a great thick book on evolutionary
biology.

\textbf{Victor:} In a recent article (Dyson, \citeyear{dyson}),  Freeman Dyson
classifies mathematicians as frogs and birds; or as Erich Lehmann put it
(Lehman, \citeyear{lehmann}): problem solvers and system builders. Where do you stand?

\textbf{David:} I like to be a bit of both. I~like solving problems,
but yet from my math background\vadjust{\goodbreak} I like to abstract things. I~like to transfer
information between fields. So, I~have worked at the same time with a
seismologist, Bruce Bolt, and with a neuroscientist, Walter Freeman. Walter
works with EEG (electroencephalogram) analysis. I~would be telling Walter some
of the clever things the seismologists were doing and I would be telling Bruce
some of the clever things that the neuroscientists were doing. They each
could then be thinking of applying these things to their own data. Abstraction
was the route between the two fields. Transfer of knowledge is a~topical goal
and the politicians like it a~lot. It probably makes sense because you can
``start sooner'' in a different field. Dyson by the way is another hero. I~think
I read various of his books and papers. I~used to look a lot at the physics
literature.

\textbf{Victor:} Do you have a favorite paper?

\textbf{David:} I believe that my favorite papers are the ones that I
had to work the hardest to get the result. I~believe I told you I had solved all
the problems, except one, in Sam Wilks' book. The one which was about getting an
asymptotic joint distribution of the median and the mean. I~did not know how to
get that and when I told Sam I don't think he knew how either. He said he had
found the result in a paper by some Hungarians. I~never found that paper either.
Eventually, I~ran into the notions of strong approximations, later called
coupling, and read a report by Ron Pyke---another role model of mine---and one
of his students, on getting a strong approximation for the empirical CDF using
tied down Brownian motion. But for the problem I was concerned with, I~needed an
error term. I~think I was the first to set down that approximation with an error
term. The Hungarians then referred to my work and generalized it to get a lot of
wonderful results.

\textbf{Victor:} You're referring to your early \textit{Bulletin of the AMS}
paper on the representation of an empirical distribution function (Brillinger, \citeyear{brillempir})?

\textbf{David:} That's right. That's one of my
favorites. It just opened up a whole host of things. Then, of course, when you get
such a result you can improve it a great deal. But this strong approximation
just lets you write down results using standard calculus. That was an important
one to me.

\textbf{Victor:}  And what about a ``favorite rejected paper,'' or, to
put it differently, is there an instance when you might have felt angry at a referee?

\textbf{David:} No, never anger at an academic referee, sometimes anger
at a soccer referee (Victor laughs). I~had a paper once, that I
thought was quite interesting, on a representation for polymeasures.\vadjust{\goodbreak} So
polymeasures do relate to polyspectra, but really it was more useful for nonlinear operators. I~mean there's this huge
world of linear operators, but polymeasures provide you with representations for
an important class of polynomial operators. And then, since I was just about to
move to England, I~thought it would make sense to send it to the \textit{Journal of the
London Mathematical Society}. To this day, I~think that if I had actually been at
LSE and sent it from there, they would have accepted it. But I~just got
a~referee's report back saying that they were just not interested in that type of
paper. I~was young, I~was learning. I~still had the attitude that I'd rather be
playing hockey than doing this stuff, and that stood me a good stead. Really,
that's not made up. Plus, I~had Tukey telling me that he had many papers
rejected. I~think I read somewhere that Rob Tibshirani said that his first ten
papers were rejected. Tukey's thing was resubmit somewhere else. I~sent it to
the \textit{Proceedings of the American Mathematical Society} and they accepted it
directly (Brillinger, \citeyear{brillprocAMS}).

Tukey and I had a paper rejected by two journals (Brillinger and Tukey, \citeyear{brilltukey}). He
told me not to worry, it could appear in his \textit{Collected Works}, and it did.

\textbf{Victor:} Going in the other direction, was there a~paper that
you found had much more impact than what you would have expected?

\textbf{David:} I just love to do math problems. All through High
School and University, there were problems from the \textit{American Mathematical
Monthly} that I would try to solve. So, I~was doing it for my amusement. You know,
you could send a solution and sometimes they would publish it. So, I~think in
many cases that's why I~was doing things: there was a~problem, and I was there.
So, the polyspectra paper (Brillinger, \citeyear{brillpolyspectra}) just started out from
having fun. I~found that cumulants were a way to go. They had this property that,
if there was a multivariate variable, and if some set of its variables was
independent of the rest, then the joint cumulant was zero. This takes one
directly to a definition of mixing for general stationary processes. Perhaps the
Russians knew that result, but anyway. But I was working on this for
fun. At one point, Tukey mentioned the word, polyspectra, and I made the
connection---and wrote that paper. That paper might have helped me get some
invitations to speak and job offers and promotions. It surely led to my
collaborating with Murray Rosenblatt.

\textbf{Victor:} Well, it's been cited over 200 times, I~think!

\textbf{David:} I remember I gave a talk on that research at Cambridge.
David Kendall, whose work you know well, had invited me. When I was done with
the talk, I~think he was as baffled as most other people were by what I was up
to. Maybe I was just not good at explaining it. Hopefully, I~eventually learned
how to do so. Anyway, Kendall said something like, ``Now let's go have some poly-tea in
our poly-cups.'' So that broke the ice (laughs). Most of these great
people have a sense of humor. They can seem pretty serious because one has to
think hard to do the research. But you realize that basically they're people who
have families, and have fun with their children at the playground. There is a
human side to all of them. So, in the beginning, very few people would refer to
that paper at all. I~think Kolmogorov knew about it, and I had a bit of an
interaction with Zurbenko about it. But that was pretty much it. But then, in the
early '80s all of a sudden I get this flood of reprint requests! This was when
people still used reprints, they didn't have things on the web. And so, all of a
sudden I'm being invited to these conferences, some of them in exotic places, on
``Higher Order Spectra''---that's what they called it. My preference is
cumulant spectra. I~remember saying things at some of these conferences, like,
``Nothing matters unless you show it used on a real data set.'' And I remember
seeing some of the engineers looking at each other. Because in so many cases
they would tend to use proof by simulation. That gave them the feeling they had
done their duty in terms of a proof. I~don't put them down, I~have a huge amount
of respect for engineers. My favorite committees are engineering committees
because they have something better to do than being on the committees! And they
have this attitude, that Allin Cornell, an earthquake engineer expressed
to me once, the attitude that every engineering problem has a solution. And
I~think Tukey was showing me that many times over in the form that every
statistics problem has a solution. And that it's the statistician's
responsibility to find it. You can't just abandon a scientist and their data.

\textbf{Victor:} On your office door in Evans Hall there is a sticker:
$2\pi\neq 1$. Would you care to elaborate on this for the uninitiated?

\textbf{David:} Oh well, yes, that's my logo! I usually like to make
people figure it out. It goes back a long way. Here's one story: this student,
Raffa (Irizarry) whom I~have mentioned already, was just a joy. I~would hear
loud footsteps of someone running down the corridor toward my office. And then
Raffa would appear, slide me off my chair, and open a window on my computer
saying, ``You have got to see this!'' One day he ran into my office saying, ``I
found it! $2\pi$ is not 1!'' He had discovered what was going wrong in his
computations by simulating the basic procedure countless times for a known case.
His answer was out by a multiple of $2\pi$. Raffa was already a~modern
statistician using Mathematica and simulation to deal with analytic problems. By
the way, he just received COPSS' Young Statisticians Award. That made me very
proud. Peter Guttorp just got an honorary degree from his home University of
Lund. The grad students have been my great joy at Berkeley. Ross Ihaka received
the Pickering Medal in New Zealand for his work in developing the statistical
package \texttt{R}. Others too. I~mean my students make me proud for their research and
professional contributions. John Rice has excelled in those two areas and just
completed a second successful term as our Department Chair. They are grandchildren of Tukey's, and a lot of what they
are getting from me is what I learned from Tukey. For example, you've seen me
filing papers with these plastic ziplock bags? Well this is a Tukey idea from
many years ago! Victor, does Stephan (Morgenthaler) ever do that?

\textbf{Victor:} I don't recall, I'll make sure to check!

\textbf{David:} Well, you can tease him about it. If he says no, tell
him that Brillinger says he would have a better career using these bags! He will
have an answer to that, I'm sure
(both laugh)!

\textbf{Victor:} Churchill (Churchill, \citeyear{churchill}, p. 17) wrote something
like, ``All students should learn English, and then the clever ones should take
Latin as an honour and Greek as a treat.'' Translated into mathematical or
statistical topics, what would be your pick?

\textbf{David:} You could probably ask me that five times and get five
totally different answers! Because right now I~think it's puzzles. As a
youngster, I~was always doing problems in the newspaper, you know ``three men
are in a room and they can't see what's on their own head$\ldots$'' and things like
that. I~had a lot of fun in doing that and a lot of good intellectual exercise.
Perhaps the exercises in my book was the part I enjoyed most. It was
the hardest part too. The things I had to work hardest on are the ones I respect
the most. I~developed an estimation method and a paper once, on my bike ride
home. I~had the idea, went to the typewriter upstairs, sat down, and typed it
up. I~sent it to\vadjust{\goodbreak} \textit{Biological Cybernetics} directly (Brillinger, \citeyear{brillER}).  All done
in a couple of hours! That didn't impress me. Then, there are some other things
like how to handle the ``integrate and fire'' model in neuroscience
(Brillinger and Segundo, \citeyear{brillsegundo}), which took quite a while to come along.

\textbf{Victor:} As we already mentioned, you will have supervised 40
Ph.D. dissertations by next January. What would be your advice to the next
generation?

\textbf{David:} It seems to me that learning mathematics is nowadays
being replaced by learning computer science. I~think it would be good for
students to learn near equal amounts of each of these. Computer
science  lets one check out proposed methods, learn about data structures---after all the
data are typically in a computer---and get approximate answers. But
I am not sure it really takes you to the essence of a lot of situations. Think
of the neural net models. They can be justified by the science, as in the
threshold case mentioned above. However, I~am uneasy about throwing everything in
there and getting an answer without a scientific interpretation. I~would rather
use something that has scientifically interpretable parameters. Let me add,
though, that I am certainly not averse to using some tool to see what it can do
for me. I~would like to see students come back to studying more serious
mathematics. I'm astonished that some students in the computer science community
don't know elementary trigonometric identities.  For them, the Fourier transform
is just the FFT: you put this in and you get this out. People learn a lot by
just doing something and seeing what you get. That's a system identification
approach where one inputs a signal and sees what comes out. I~think it is a lot
more rewarding to really get some understanding of \textit{why} it is happening.
Although in science it doesn't always work that way. I~remember Fred Mosteller
saying many years ago that nobody knew then why aspirin worked, but that of
course we are going to use it because it appeared to work.  But still I~think
learning what the thing was doing is fundamental, because then you can improve
on it.

My bottom line is: have fun! That sounds trite but I'm serious. If you are
worried about something, consider what you can do about it. If there is
something, do it. If not, what's the point of worrying? When you have a child
die after a very long battle with cancer, as Lorie and I did, you simplify a lot
of things. You take things to their essence. Don't be afraid to cry. It is
another thing you learn going through a tragedy. Many say crying is hard
sometimes. For me, it just happens.

\textbf{Victor:} David, thank you very much for sharing these memories
of your remarkable life and career. But I have to ask one last question: would
you still rather have been a hockey player?

\textbf{David:} Oh yes!!! (laughs out loud)  There is noooooo doubt in
that! I gave the after-dinner talk at one of the Canadian Statistical Society
meetings and the title was: ``Why I became a Statistician.'' You can guess what
the punch line was!

\textbf{Victor:} Thanks again, David.

\textbf{David:} Thank you, Victor. You had some good questions. I~mentioned only some of
my students. I~probably have an anecdote about each, but
I'll save those for another time.

\section*{Acknowledgments}
We would both like to thank Ms. Maroussia Schaf\-fner for her patient work in
transcribing our nearly four-hour conversation. Parts have been lightly
edi\-ted and reorganized for clarity. David thanks the many people who
have helped him during his career.

\vspace*{-1pt}

\end{document}